\documentclass[fleqn,usenatbib,useAMS]{mnras}

\usepackage{graphicx}	
\usepackage{amsmath}	
\usepackage{amssymb}	
\usepackage{multicol}        
\usepackage{bm}		
\usepackage{pdflscape}	
\usepackage[nameinlink,capitalise]{cleveref}

\newcommand{\mbh}{ M_{\rm BH}}

\newcommand{\Msun}{\rm M_{\sun}}

\newcommand{\kpc}{\rm kpc}

\newcommand{\ckpc}{\rm ckpc}
\newcommand{\pkpc}{\rm pkpc}
\newcommand{\cMpc}{\rm cMpc}
\newcommand{\Mpc}{\rm Mpc}

\newcommand{\lsim}{\mathrel{\hbox{\rlap{\lower.55ex\hbox{$\sim$}} \kern-.3em\raise.4ex\hbox{$<$}}}}
\newcommand{\gsim}{\mathrel{\hbox{\rlap{\lower.55ex\hbox{$\sim$}} \kern-.3em\raise.4ex\hbox{$>$}}}}

\newcommand{\Rvoid}{r_{\rm void} }

\newcommand{\tform}{t_{\rm form,*}}
\newcommand{\EAGLE}{\textsc{eagle} }
\defcitealias{schaye2015}{S15}
\defcitealias{crain2015}{C15}

\usepackage[T1]{fontenc}
\usepackage{ae,aecompl}

\usepackage{newtxtext,newtxmath}
\usepackage{ulem}

\title[Void Galaxies]{Revealing the properties of void galaxies and their assembly using the EAGLE simulation}

\author[Rosas-Guevara et al.]{Yetli Rosas-Guevara$^{1}$ \thanks{E-mail:yetli.rosas@dipc.org}, Patricia Tissera$^{2,3}$, Claudia del P. Lagos$^{4,5}$, Enrique Paillas$^{6,7}$, Nelson Padilla$^{8}$ \\
$^{1}$ Donostia International Physics Centre (DIPC), Paseo Manuel de Lardizabal 4, 20018 Donostia-San Sebastian, Spain\\
$^{2}$Institute of Astronomy, Pontificia Universidad Católica de Chile, Santiago, Chile.\\
$^{3}$Centro de Astro-Ingenier{\'i}a, Pontificia Universidad Católica de Chile, Santiago, Chile \\
$^{4}$International Centre for Radio Astronomy Research (ICRAR), M468, University of Western Australia, 35 Stirling Hwy, Crawley, WA 6009, Australia.\\
$^{5}$ARC of Excellence for All Sky Astrophysics in 3 Dimensions (ASTRO 3D) \\
$^{6}$Waterloo Centre for Astrophysics, University of Waterloo, Waterloo, ON N2L 3G1, Canada.\\
$^{7}$Department of Physics and Astronomy, University of Waterloo, Waterloo, ON N2L 3G1, Canada.\\
$^{8}$Instituto de Astronom\'\i a Te\'orica y Experimental (IATE), Comisi\'on Nacional de Investigaciones Cient\'\i ficas y T\'ecnicas (CONICET), Universidad Nacional de \\ C\'ordoba, Laprida 854, X500BGR, C\'ordoba, Argentina. \\}
\date{Last updated 2020 June 10; in original form 2013 September 5}

\pubyear{2022}

\begin{document}
\label{firstpage}
\pagerange{\pageref{firstpage}--\pageref{lastpage}}
\maketitle

\begin{abstract}
 We explore the properties of central galaxies living in voids using the \EAGLE cosmological hydrodynamic simulations. Based on the minimum void-centric distance, we define four galaxy samples: inner void, outer void, wall, and skeleton. We find that inner void galaxies with host halo masses $<10^{12}\Msun$ have lower stellar mass and stellar mass fractions than those in denser environments, and the fraction of galaxies with star formation (SF) activity and atomic hydrogen (HI) gas decreases with increasing void-centric distance, in agreement with observations. To mitigate the influence of stellar (halo) mass, we compare inner void galaxies to subsamples of fixed stellar (halo) mass. Compared to denser environments, inner void galaxies with  $M_{*}= 10^{[9.0-9.5]}\Msun$ have comparable SF activity and HI gas fractions, but the lowest quenched galaxy fraction.  Inner void galaxies with $M_{*}= 10^{[9.5-10.5]}\Msun$ have the lowest HI gas fraction, the highest quenched fraction and the lowest gas metallicities. On the other hand, inner void galaxies with $M_{*}>10^{10.5}\Msun$ have comparable SF activity and HI gas fractions to their analogues in denser environments. They retain the highest metallicity gas that might be linked to physical processes that act with lower efficiency in underdense regions such as AGN feedback. Furthermore, inner void galaxies have the lowest fraction of positive gas-phase metallicity gradients, which are typically associated with external processes or feedback events, suggesting they have more quiet merger histories than galaxies in denser environments. Our findings shed light on how galaxies are influenced by their large-scale environment.
\end{abstract}



\section{Introduction}

From large galaxy surveys, it is well known that the large-scale structure can be described as a 3D cosmic web with thin filaments connected by galaxy clusters and sheets that surround underdense regions \citep{bond1996}. The current cosmological paradigm, $\Lambda$-CDM, explains the cosmic web origin as the product of the formation and evolution of primordial density perturbations superposed on a homogeneous and isotropic background. The underdense regions, known as cosmic voids, account for more than half of the entire volume of the Universe \citep{dacosta1988,pan2012} with sizes (diameters) that lie between $\sim 1$ to $\sim 100$ Mpc $h^{-1}$. Due to their sensitivity to some cosmological constraints, such as dark energy \citep[e.g.][]{li2011,cai2015,pisani2015}, cosmic voids have recently been exploited for cosmological tests \citep{paillas2021}.

Cosmic voids are especially important because they could be used as ideal settings to investigate the influence of the large-scale environment on the formation and evolution of galaxies since they are expected to be less evolved and retain the memory of a more primitive Universe \citep[see review by][]{vandeweygaert2011}. The formation and evolution of galaxies entail incredibly intricate, interconnected, and multiscale processes such as galaxy mergers and tidal effects, are included. It is anticipated that galaxies inhabiting cosmic voids are primarily assembled by internal processes, and that their features will differ from those inhabiting other environments.
Indeed, many observational studies employing galaxy surveys have identified differences in some properties between void galaxies and those residing in denser regions \citep[e.g.][]
{szomoru1996b,rojas2004,hoyle2012,kreckel2012,florez2021}. Generally, void galaxies contain less stellar mass \citep [e.g.] []{croton2005,moorman2015}, are bluer \citep [e.g.] []{grogin2000,rojas2004,padilla2010,hoyle2012} and with later-type morphology \citep [e.g.] []{rojas2004,croton2005}. There is no consensus regarding the differences  between  some properties of void galaxies and those in denser environments  with comparable stellar mass. For instance, using a huge sample of galaxies from SDSS-DR7, \cite{ricciardelli2014}, classifying voids as large spherical regions devoid of galaxies ($\gsim 10\, \Mpc\,\rm h^{-1}$) and shell galaxies as those galaxies located at a distance $\geq 30\, \Mpc\,\rm h^{-1}$ from the centre of a void, have found that void galaxies had higher star formation activity than  those lying  in the shell of the void or a control galaxy sample. In contrast, when only star forming galaxies were considered, they exhibit the same SF activity as shell void galaxies and a control sample with the same stellar mass distributions.
\cite{moorman2016}, utilizing an optical large galaxy sample from SDSS DR8 \citep{blanton2011} in conjunction with HI detections from the ALFALFA survey \citep{haynes2011}, found that void galaxies have similar star formation activity to wall galaxies with similar stellar mass. The authors employed a spherical void catalogue  and considered wall galaxies to be those outside of the voids. Similarly, \cite{beygu2016} using the Void Galaxy Survey (VGS), which uses a combination of the Delaunay Tessellation Field Estimator and a Watershed Void Finder  to identify a void, found similar results when comparing void galaxies with galaxies in the field that are everywhere except in void interiors. Recently, \cite{dominguez2022} found  comparable  mean values of the specific star formation rates (sSFRs) for  void and field galaxies when the sample is limited to star-forming galaxies using 20 cosmic voids that are part of the VGS. Furthermore, the authors found that the molecular and atomic gas masses in void galaxies are comparable to those in galaxies in wall and filaments. These results are in contradiction with \cite{florez2021} who found that
that void galaxies had higher atomic gas masses than galaxies in filaments and walls, using the RESOLVE survey and ECO catalogue and defined as void galaxies the 10\% of galaxies having the lowest local density.

Using spectroscopic and SSDS data from 40 dwarf galaxies in the Lynx-Cancer Void, \cite{pustilnik2011}  showed that these galaxies have lower gas-phase metallicities than those in denser settings in the Local Volume. In contrast, \cite{kreckel2015} studied dwarf galaxies carefully selected from the inner parts of seven cosmic voids and compared them to  isolated dwarf galaxies from existing samples. The authors found that the gas-phase metallicity of both samples were similar, indicating that external gas accretion could play a minor role in the chemical evolution of these systems.  However, we stress that these findings are based on small samples of galaxies. Consequently, much remains to be understood.

Gas-phase metallicity and gas fractions in galaxies could also hold information about the assembly history of void galaxies that could be connected to the properties of their host halo. In principle, because gas properties are dependent on the gravitational potential well, the ratio of stellar-to-dark matter halo mass in a galaxy should alter them. Increased depth of potential wells prevents a greater fraction of gas from escaping the galaxy due to supernova-driven winds, tidal stripping, and other mechanisms. \cite{douglass2019} estimated the stellar-to-dark matter halo mass relation for galaxies in the void and non-void regions by calculating the relative velocity of H$_{\alpha}$ emission line across the galaxy surface to measure the rotation curve of each galaxy in the MaNGa survey and identifying voids with the spherical VoidFinder of \cite{hoyle2002}. Those galaxies than are not in the void are referred to wall galaxies. The authors found no significant difference in the stellar mass-to-dark matter halo mass ratios between void galaxies and wall galaxies.

Despite the fact that significant progress has been made, all the previously described inconsistencies may be attributable to either the variances in the method used to identify a cosmic void \citep[e.g.][]{colberg2008,cauntun2018,paillas2019}, the potential bias of the tracers utilised \citep[e.g.][]{paillas2017,florez2021}, or statistical errors associated with the size of the tracer samples. Future observational surveys, such as the CAVITY project\footnote{https://cavity.caha.es}, will provide a survey of massive void galaxies that will assist in identifying the origin of these disparities.

From a theoretical point of view, whereas cosmological \textit{N-body} simulations have been widely used to characterise cosmic voids, it was not until relatively recently that cosmological hydrodynamic simulations have provided a theoretical framework for the formation and evolution of galaxies \citep{dubois2014,schaye2015,nelson2018}. In particular, large-volume cosmological hydrodynamic simulations are well suited for studying the formation and evolution of galaxies in voids. \cite{paillas2017} used the largest simulation of \textsc{eagle} \citep{schaye2015} to identify and characterise voids, as well as, the impact of baryons on their distribution and properties (e.g., void size and void density profiles).  The authors concluded that, overall, baryons have no discernible effect on void statistics. \cite{alfaro2020} shows that the halo occupation distribution in cosmic voids cound be different than in other large-scale structures using the \textsc{TNG300} simulation \citep{nelson2018} and identifying voids via Voronoi tessellation of the galaxy catalogues \citep{ruiz2015a}.
\cite{habouzit2020} study the galaxy properties and their black hole properties in voids identified in the \textsc{HorizonAGN} simulation \citep{dubois2014}. The authors use two distinct methods to identify void structures and find that low-stellar mass galaxies with intense star-forming activity are more frequent in the inner regions of voids, whereas their black holes and their host galaxy grow together in voids in a similar way to denser environment, even though the growth channels of BHs and their host galaxies in cosmic voids are different from denser environments \citep{ceccarelli2022}.

In this paper, we aim to conduct a systematic investigation of the properties of central galaxies as a function of their location on the cosmic web using the largest simulation of the \EAGLE project. This simulation, although it is relatively small compared to the largest voids observed in the Universe, allows us to have a representative number of galaxies in diverse regions. This analysis is a step forward to the next generation of cosmological simulations, which will be focused on larger volumes. The \textsc{eagle} simulations reproduce the low-redshift properties of a galaxy population in agreement with observations \citep{schaye2015,crain2015,mcAlpine2016}. \cite{lagos2015} find a remarkable agreement between the H$_2$ properties of \EAGLE galaxies and those that are observed. \cite{tissera2019} demonstrated that the \EAGLE simulation is able to reproduce observed the large diversity of gas-phase metallicity gradients. Investigating the highest resolution \EAGLE simulation, \cite{tissera2021} analysed the evolution of the gas-phase metallicity gradients, highlighting the importance of mergers in driving the diversity of metallicity gradients observed today.

We use the void catalogue developed by \cite{paillas2017}, which is based on a spherical underdensity finder that characterises the voids by their centre position and size. We construct four galaxy samples based on the void-centric distance, which is defined as the distance to the centre of the nearest void for a given galaxy, and consequently investigate the properties of galaxies regarding the large-scale environment. To clearly distinguish mass dependence from environmental dependency (i.e., the effects driven by the overabundance of low-stellar mass galaxies in voids), we additionally investigate subsamples of galaxies with identical stellar mass distributions. It is important to note that our goal is not to do a direct and fair comparison with each observational dataset, since we would need to account for all of the many systematic elements that may possibly bias our comparison with a particular observational data. However, in order to guide the reader, we cite global tendencies identified via observations.

The outline of the paper is organised as follows. We describe the \EAGLE simulations, the void finder, and the definitions of the parent galaxy samples according to the void-centric distance in section \ref{sec:data}.  In Section \ref{sec:galwithdist}, we investigate the main properties of galaxies as a function of their distance to the nearest void. In Section \ref{sec:galprop}, we investigate the galaxy properties for galaxy subsamples with identical stellar mass distribution as galaxies located in the internal parts of the voids.  We then explore the evolution and assembly of galaxies using the subsamples in section \ref{sec:assembly}. Finally,  we discuss and summarise our findings in Sections \ref{sec:discussion} \& \ref{sec:summary} respectively.

\section{Methodology}
\label{sec:data}
\subsection{ Simulations}

\begin{figure*}
	\includegraphics[width=2.\columnwidth]{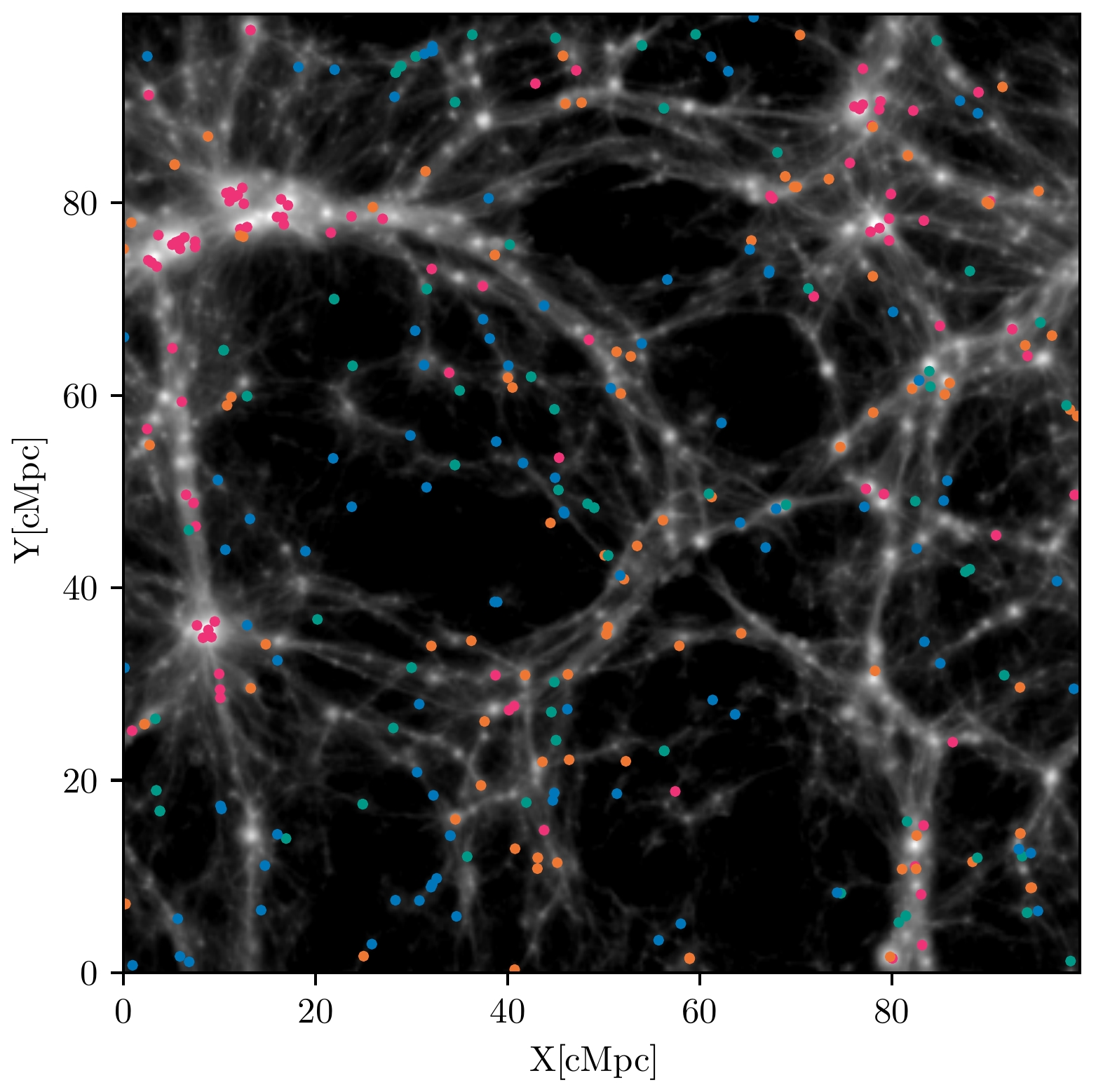}
    \caption{A slice of $100\times100\times 25\,\cMpc$  of galaxies in the \EAGLE largest simulation at $z=0$. With the same galaxy number density, blue and green circles correspond to inner void and outer void galaxies, respectively. Orange and magenta circles represent wall and skeleton galaxies, respectively.}
     \label{fig:xydiagram}
\end{figure*}

We use the largest simulation of the \EAGLE project\footnote{http://eaglesim.org \\
http://eagle.strw.leidenuniv.nl}  \citep{schaye2015,crain2015}, which consists of  a suite of cosmological simulations with varying galaxy formation subgrid models, numerical resolutions and volumes.  The simulations were performed with a modified version of the SPH code P-Gadget 3 that is an improved version of  Gadget 2  \citep{springel2005b}. This code includes galaxy formation subgrid models to capture unresolved physics including cooling, metal enrichment and energy input from star formation \citep{schaye2008} and black hole growth \citep{rosas-guevara2015}.  A full description of  the \EAGLE project is found in \cite{schaye2015} \& \cite{crain2015}.
The largest simulation of \EAGLE has a comoving volume of  $(100\, \cMpc)^3$ with  a mass resolution of $9.7 \times  10^6 \,\Msun$ for dark matter (and $1.81 \times 10^6 \Msun $ for baryonic) particles and   a softening length of $2.66 \,\ckpc$ \footnote{Throughout the paper, we refer  comoving distances  by preceding a 'c' in $\kpc$  and physical lengths  by  a 'p' as  \pkpc.}  limited to a maximum physical  size of $0.70\, \pkpc$.  The simulation adopts the $\Lambda$-CDM cosmology from \cite{planck13} with cosmological parameters: $\Omega_\Lambda=0.693$, $\Omega_{\rm m}=0.307$, $\Omega_{\rm b}=0.04825$, $\sigma_8=0.8288$, $h=0.6777$, $n_{s}=0.9611$ and $Y=0.248$ (see Table 1 from \citealt{schaye2015} for details).

The simulation outputs were analysed using the SUBFIND programme
to identify bound control structures \citep{springel2001,dolag2009} within each Friends of Friends (FOF) dark matter halo. These substructures are identified as galaxies and measure stellar masses within a radius of 30 $\pkpc$ \citep{mcAlpine2016}.  FOF dark matter halo masses, $M_{\rm halo}$, are defined as all matter within the radius $r_{200}$ where the mean internal density is $200$ times the critical density.
We consider the ‘central’ galaxy as the galaxy closest
to the centre (minimum of the potential) for each FOF structure. The remaining galaxies within the FOF haloes are classified as its satellites.

\subsection{Void Catalogue}

We employ the void catalogue presented in \cite*{paillas2017} at $z=0$. Here, we provide only  a brief description of the catalogues; an extensive description of the void identification and associated void statistics in EAGLE are found in \cite*{paillas2017}. A spherical underdensity finder based on the algorithm described in \cite*{padilla2005} is used to locate cosmic voids. The algorithm begins by building a rectangular spatial grid and counting the number of galaxies within each grid cell. The centres of empty cells are considered candidates for void centres.  Around each candidate, spheres are grown until the integrated galaxy number density in the sphere surpasses $20$ per cent of the mean galaxy number density.  The void radius, $\Rvoid$, is defined as the radius of the largest sphere satisfying this criterion surrounding a given centre. If two adjacent voids have centres that are closer to a set percentage of the sum of their radii, the smallest of two is discarded from the catalogue. To validate $\Rvoid$ for the remaining voids, the void centre is moved in various directions, and if the new radius is larger than $\Rvoid$, the position of the void centre is updated. \cite{paillas2017} employed several tracers to identified voids and as well as  varying degrees of  overlap to examine the implications on voids statistics. In this study, we use the void catalogue in which galaxies with a stellar mass $\geq 10^8\Msun$ are  tracers and there is  $40$ percent overlap in the extent of the voids.  The cumulative distribution of void sizes are  depicted in Fig. 6  and  Table 1 of \cite{paillas2017} for different tracers. Particularly, for the catalogue used in this work, there are $709$ voids whose sizes range from $4.9$ to $24.3$ with an mean of $7.0$ pMpc. Approximately $10$ per cent of the voids are larger than $10$ pMpc and $1$ per cent of the voids are larger than $15$ pMpc.

\subsection{Morphology and gas-phase oxygen abundances gradients}
\label{subsec:morphclass}
The method described in \citet{tissera2012} is used to determine the morphological classification of galaxies. Galaxies are first rotated so that the z-axis is located in the direction of the total angular momentum of the stellar component.
Then,  using the circularity parameter $ \epsilon = J_{\rm z}/J_{\rm z,max} (E)$,
 they are divided into a stellar disc and a bulge component, where $J_{\rm z}$ denotes the angular momentum and $J_{\rm z,max} (E)$ is the maximum $J_{\rm z}$ over all stellar particles at a given binding energy $E$. The disc component is defined as those stellar particles with, $ \epsilon >0.5$  while the bulge components comprise those
with smaller $\epsilon$ and are more gravitationally bounded than $E$ evaluated at $0.5 R_{\rm hm}$ (being $R_{\rm hm}$ the half mass radius).
The disc-to-total ($D/T$) ratio is defined by the ratio of the mass of the disc to the total stellar mass,  where the total mass is defined as the sum of the disc and bulge masses such as  $B/T+D/T=1$. We select, as disc-dominated galaxies,  those with $D/T\geq0.5$.

The radial gradients of gas-phase oxygen abundances were taken from \citet{tissera2019}, for star-forming galaxies having a disc component containing more than 1000 baryonic particles and more than 100 star-forming gas particles\footnote{These number have been selected to diminish numerical noise due to low number of particles in the determination of the metallicity profiles. This implies that the catalogue at $z=0$ includes galaxies with $M_{*}> 10^9 \Msun$ which have star formation rates within the range $[0.5-20]\Msun/\rm yr$, approximately \cite{tissera2019}. }. This condition ensures to capture the metal distribution from the star-forming gas phase.The selected galaxies have a variety of morphologies. The radial metallicity profiles are estimated by weighting the oxygen abundances of the
star-forming gas particles by their star formation rate. This improved
the comparability of the simulated gradients to observations. Finally, the metallicity gradients are estimated using a linear regression fit inside the radial range $[0.5, 2]R_{\rm hm}$.

\subsection{Parent-galaxy samples}
\label{subsection:galaxysample}
Our study is limited to galaxies with a stellar mass higher than $10^9\Msun$within a-$30$-$\pkpc$ spherical aperture\footnote{It should be noted that this is a higher stellar mass cut than the one used to identify voids. The reason for higher stellar mass cut in this study is to ensure that the properties and evolution of the galaxies are well captured. As a consequence, the number of galaxies are the same order of magnitude as the number of voids.}. We determine the distance between the centre of each galaxy, defined as the position of the minimum potential well, and the centre of all voids provided by the void catalogue. Each galaxy is then paired with the void whose centre is the closest to the centre of the galaxy. This ensures that each galaxy is paired a unique void. This distance will henceforth be referred to as the void-centric distance.
To determine the environmental effects of the large-scale structure on the galaxies, we consider the shape of the void density profiles, as shown in  Fig.~7 of \citet{paillas2017}. The density of the voids in their inner regions is nearly constant and much smaller than the mean galaxy number density. As we approach to the void boundaries, the density rises steeply, peaking at the void radius and converging to the mean on larger scales. Taking this into account, we split the selected galaxies into four samples based on their void-centric distance and in terms of the void radius:
\begin{itemize}
    \item {\bf Inner void:} galaxies are defined as those whose minimum void-centric distance is between $0$ and $0.8\Rvoid$, where $\Rvoid$ is the radius of the closest void.
    \item {\bf Outer void:} galaxies whose minimum void-centric distance is between $0.8\Rvoid$ and $\Rvoid$.
    \item {\bf Wall:} galaxies are those located  between $\Rvoid$ and $1.4\Rvoid$.
    \item {\bf Skeleton:} galaxies are those located beyond $1.4\,\Rvoid$.

\end{itemize}

In total, we identified $513$ inner void galaxies, $588$ outer void galaxies, $7723$ wall and $4376$ skeleton galaxies, of which  $492$, $528$, $4597$ and $1783$  are centrals, respectively. This results in  7400 central galaxies in total. Notably, each sample contains a different fraction of satellite galaxies, with the skeleton sample having a  satellite fraction of  $0.63$ whereas $0.05$ of the void galaxies are satellites. Because of this significant variation in the fraction of satellites, hereafter, we focus exclusively on the central galaxies in each sample.
In sections \ref{sec:galprop} and \ref{sec:assembly}, we select smaller samples by requesting to match the same stellar mass distribution of outer, wall and skeleton galaxies to the one of the central galaxies in the inner voids. Hence, the resulting subsamples comprise $492$ for inner voids, $461$ for outer voids, $3082$ for walls, and $1208$ for skeleton galaxies.

To visually inspect our classification, Fig.~\ref{fig:xydiagram} depicts the dark matter density map of a slice of 100x100x25 $\cMpc$ from the \EAGLE simulation. Circles represent our selected galaxies, and colours correspond to different samples.  As seen in the figure and by construction,  inner void galaxies (blue circles) are found in the less dense regions of the simulation,  whereas skeleton galaxies (magenta circles) are found in the highest-density regions. In intermediate density zones,  outer void (green circles) and wall galaxies (orange circles) are located.

\section{Galaxy properties as a function of the void-centric distance}
\label{sec:galwithdist}
\begin{figure}
    \begin{tabular}{c}
    \includegraphics[width=\columnwidth]{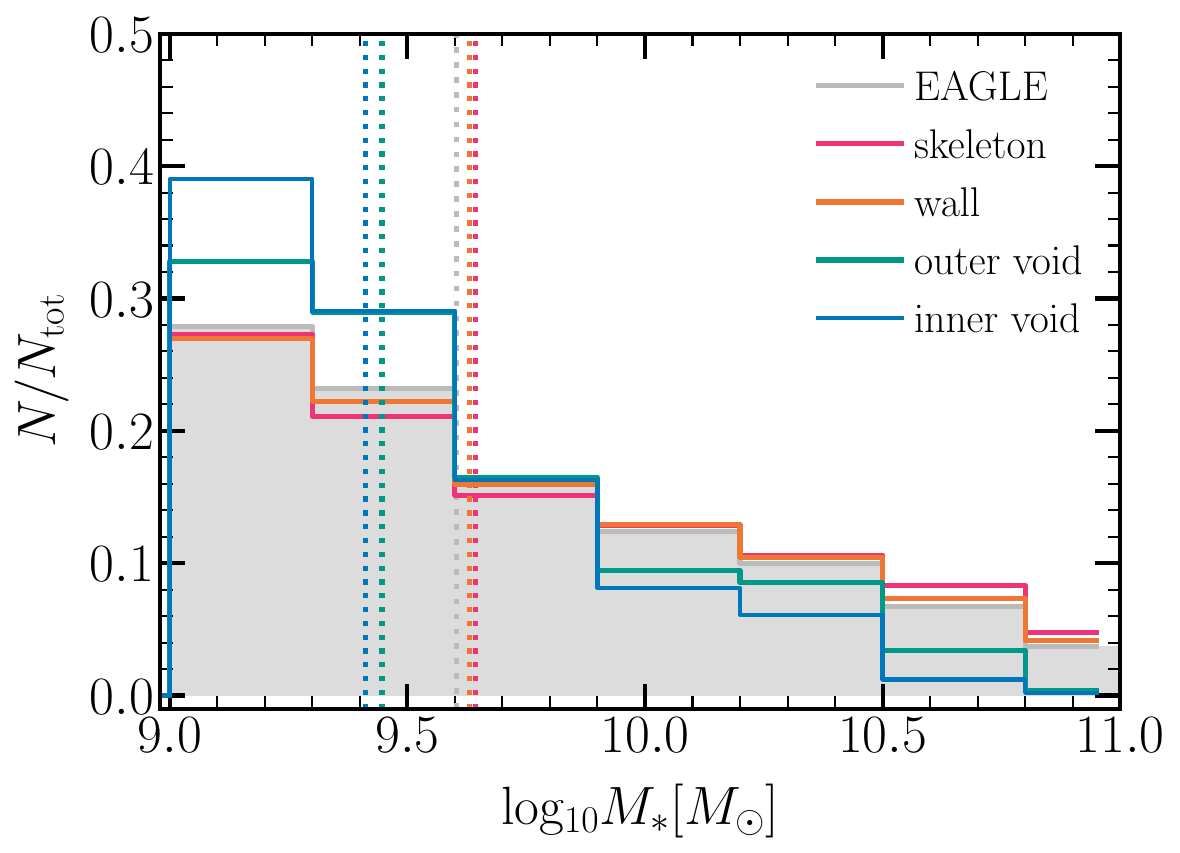} \\
    \includegraphics[width=\columnwidth]{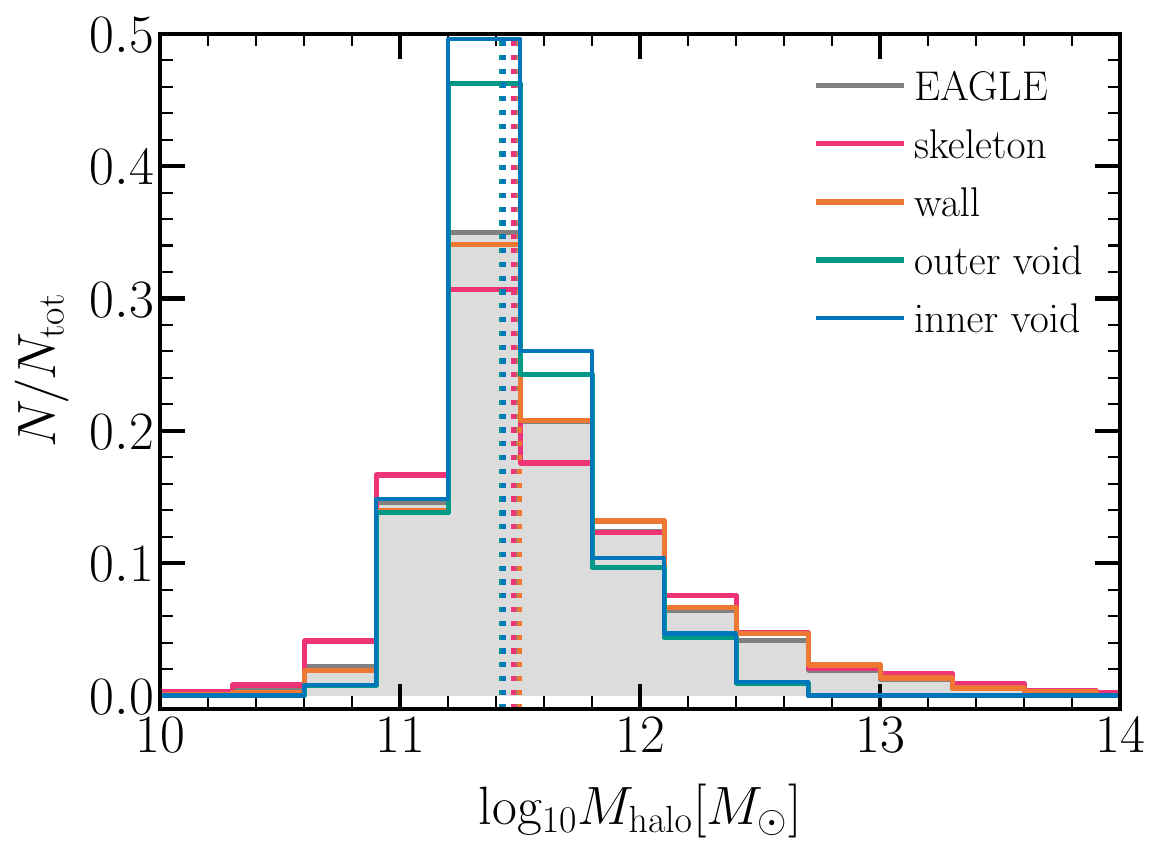}\\
    \includegraphics[width=\columnwidth]{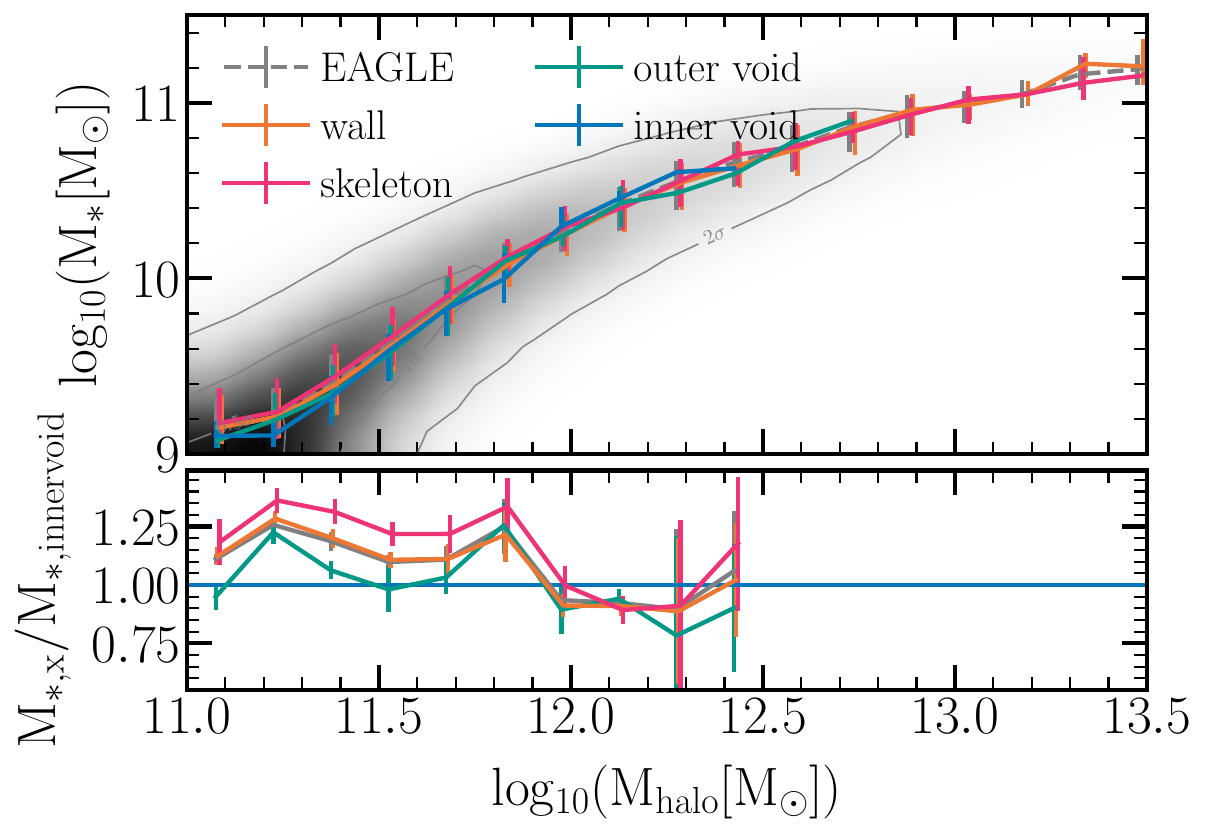}
    \end{tabular}
      \caption{ \textit{Top and middle panels:} The stellar mass and halo mass  distributions for each parent galaxy sample and the \EAGLE simulation, as specified in the legend. Vertical lines represent the median of each distribution. Inner void galaxies are biased towards low stellar mass galaxies in comparison to other regions. \textit{Bottom panel:} The median halo mass-stellar mass relation. The diffused density map and contours represent the distribution of all the central galaxies in the simulation. The error bars represent the $20^{\rm th}$ and $80^{\rm th}$ percentiles of each sample. The ratio between the stellar mass of each parent sample to the inner void sample for a given halo mass is in the bottom plot with errorbars corresponding to jackknife errors. Haloes in the inner void regions tend to host galaxies with lower stellar mass than their halo analogues from other regions for a given halo mass of $10^{[11-12.2]}\Msun$.}
    \label{fig:gmf}
\end{figure}

\begin{figure*}
	\begin{tabular}{cc}
	\includegraphics[width=\columnwidth]{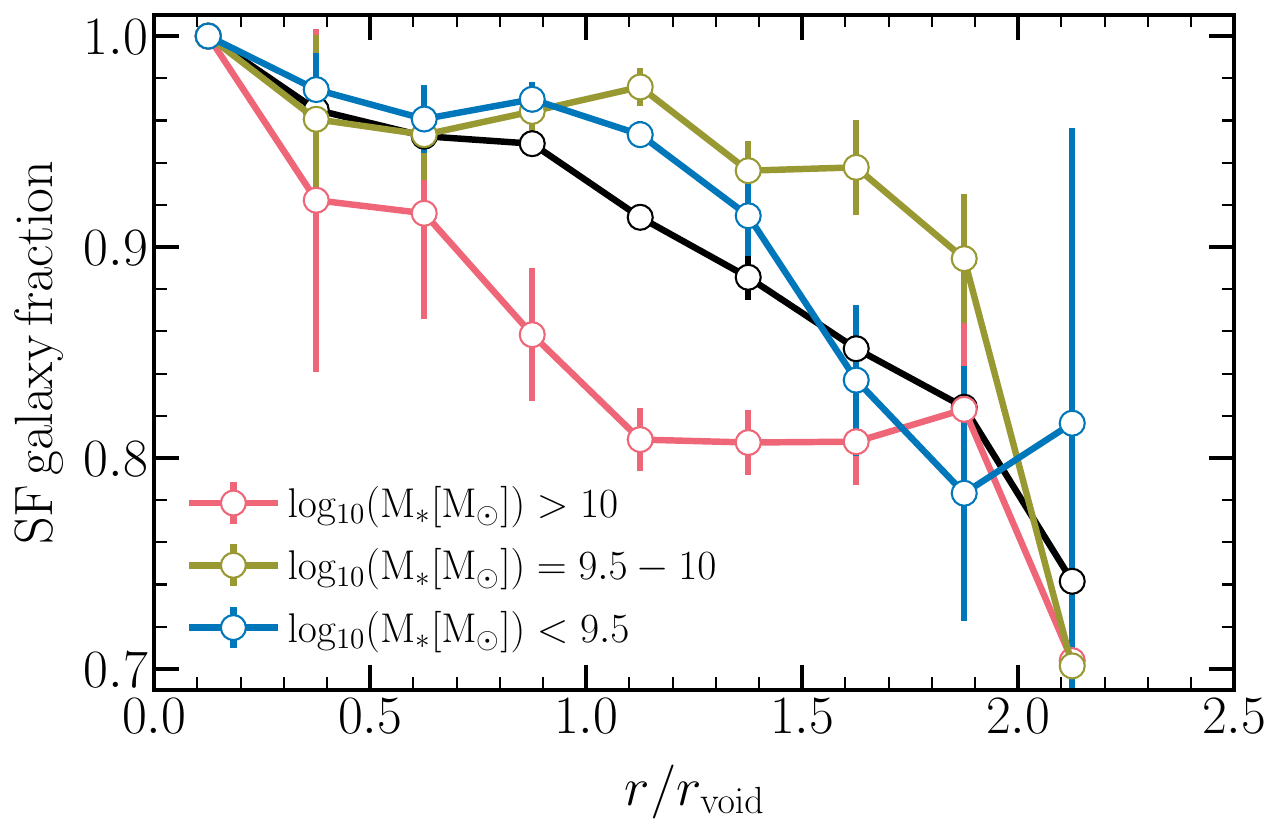} &
	\includegraphics[width=\columnwidth]{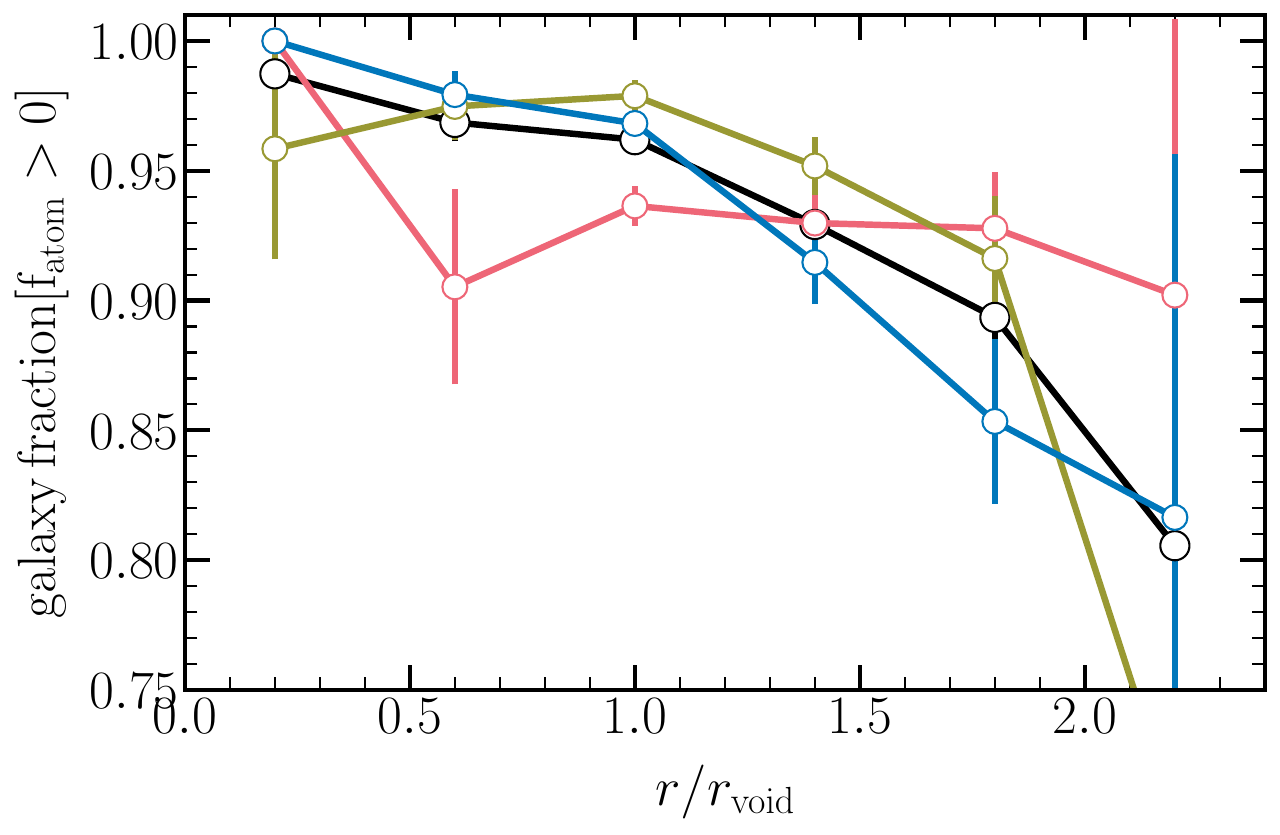}\\
	\end{tabular}
   \caption{ \textit{Left panel:} The fraction of star-forming galaxies from the parent samples as a function of the void-centric distance and for various stellar mass bins as the legend specified. Star-forming galaxies are considered to have $\rm sSFR>10^{-11.5} yr^{-1}$. The fraction of star-forming galaxies decreases with increasing the void-centric distance. \textit{Right panel}: Fraction of galaxies with Hydrogen gas fraction, $f_{\rm atom}>0$, as a function of the void-centric distance, where $f_{\rm atom}= 1.35 M_{\rm HI}/ (M_{*} +1.35 (M_{\rm HI} +M_{\rm H_{2}}))$. The fraction of galaxies with HI gas also decreases with  increasing the void-centric distance. Solid errorbars to jackknife errors using $10$ subsamples.}
    \label{fig:fractionvoids}
\end{figure*}


In this section, we explore the stellar mass and halo mass distributions, as well as their relation to our four parent samples, defined by the void-centric distance described in the previous section. Additionally, we show the abundance of star-forming galaxies with a non-zero gas fraction as a function of the void-centric distance.

In the top panel of Fig.~\ref{fig:gmf}, the stellar mass distribution of each parent galaxy sample is depicted. We observe a systematic bias toward smaller stellar masses in the stellar mass distributions of inner and outer void galaxies (green and blue solid lines). We find that the median stellar mass grows with increasing distance from the void centres. It is worth noting that the median stellar mass of the entire simulation (log$_{10}(M_*[\Msun])=9.6^{-0.3}_{+0.5}$) is comparable to that of the skeleton (log$_{10}(M_*[\Msun])=9.6^{-0.4}_{+0.6}$) and wall galaxies (log$_{10}(M_*[\Msun])=9.6^{-0.3}_{+0.5}$) as they are the more populated regions. To determine whether these differences are statistically significant or not, we perform a Kolmogorov-Smirnov (KS) test between inner void galaxies and the rest of the galaxies, obtaining p-values of $<0.08$ or less. This means that galaxies in the outer, wall and skeleton parent samples are not drawn from the same population as the inner void galaxies. These differences are consistent with previous numerical and observational works \citep[e.g.][]{croton2005,moorman2015,habouzit2020,rodriguez2022}. In particular, \citet{ricciardelli2014} used cosmic voids identified in the SDSS DR7 as larger spherical regions devoid of galaxies ($\gsim 10 \,\Mpc\,\rm h^{-1}$) and shell galaxies defined as those located at a void-centric distance larger than three times the size of the void. Even though the authors used only large cosmic voids in comparison with our void catalogue, they found low dense regions are biased towards low-stellar mass galaxies. Interestingly, when we examine the halo mass distributions (middle panel of Fig.~\ref{fig:gmf}), we see  a modest difference in halo mass  as we increase the distance to the centre of the void from log$_{10}(M_{\rm halo}[\Msun])=11.4^{-0.2}_{+0.1}$ (the $25^{\rm th}-75^{\rm th}$ percentiles) in inner void haloes  to log$_{10}(M_{\rm halo}[\Msun])=11.5^{-0.2}_{+0.4}$ in skeleton haloes which is closed to the median halo mass of the entire central galaxy population (log$_{10}(M_{\rm halo}[\Msun])=11.5^{-0.2}_{+0.4}$).

The stellar mass as a function of halo mass is one of the key properties to investigate. This might provide us with information about the relation between the assembly of stellar mass and the potential well. The median halo-mass stellar mass relation for the four parent galaxy samples and all central galaxies in the \EAGLE simulation is displayed in the bottom panel of Fig.~\ref{fig:gmf}. It is clear from the figure that galaxies in voids show lower stellar mass than galaxies in denser environments for a given halo mass in the range $10^{11}\Msun$ and $10^{12}\Msun$, although the differences are well within the scatter of the relation in each subpopulation. This is confirmed in the bottom plot of the bottom panel of Fig.~\ref{fig:gmf} that shows the ratio between the median stellar mass of each parent sample and the median stellar mass of the void galaxies for a given halo mass including the jackknife errors.
The plot shows higher stellar mass at fixed halo mass for all the parent samples of denser environments for haloes with a mass $<10^{12}\Msun$.  Additionally, for more massive haloes ($\geq 10^{12}\Msun$), the stellar mass seems to modestly increase in comparison to other environments, although we have small statistics of such massive haloes in inner void regions that is reflected in the large jackknife errorbars. As depicted  in the bottom panel of Fig.~\ref{fig:gmf},  the medians of the  skeleton (magenta) and wall galaxies (orange) extend to more massive haloes and can be seen as a tail  of more massive haloes of central galaxies in the halo mass distribution of  skeleton (magenta) and wall galaxies (orange) in the middle panel.

Compared to previous numerical works, our findings are in agreement with the study of \cite{alfaro2020} who found that haloes in void regions present less stellar mass content than haloes in other regions, using \textsc{tng300} simulation \citep{springel2018}. \cite{habouzit2020} using
\textsc{HorizonAGN} simulation \citep{dubois2014} have shown that the inner void galaxies in relatively low-mass haloes with $M_{\rm halo}<10^{11}\Msun$ were, overall, less massive than other galaxies of the simulation enclosed within haloes of same mass. We remark that  it is not our intention to conduct an apple-to-apple comparison between previous numerical efforts and our parent samples, as each study uses different void identification, bias selection, and even sub-grid physics of galaxy formation. We just intend to guide the reader on the global trends.

The fact that galaxies in voids have a lower stellar mass content than galaxies in other environments at a given $M_{\rm halo}$ suggests that star-formation activity is regulated differently in galaxies in voids than in galaxies in other large-scale environments. Indeed, it is believed that galaxies in cosmic voids evolve slowly in comparison to the overall galaxy population, remaining in the star-forming sequence in the Local Universe \citep {ricciardelli2014}.
Fig.~\ref{fig:fractionvoids} shows the fraction of star-forming galaxies (left panel) and galaxies with hydrogen gas (right panel) as a function of void-centric distance and for various stellar mass bins. The error bars correspond to jackknife errors. We define star-forming galaxies as galaxies with a sSFR$\geq10^{-11.5}\rm yr^{-1}$, where sSFR$= \rm SFR/M_{*}$ is the specific star-formation rate in an aperture of 30 pkpc. We refer to the fraction of atomic gas \citep{obreschkow2016} as $f_{\rm atom}= 1.35 M_{\rm HI}/ (M_{*} +1.35 (M_{\rm HI} +M_{\rm H2}))$ where  $M_{\rm HI}$ and  $M_{\rm H2}$ are the atomic and molecular hydrogen masses, respectively. The factor $1.35$ accounts for the universal $\sim 26\%$ helium fraction in the local Universe. The mass of hydrogen gas in atomic and molecular phase were calculated by \cite{lagos2015} for each gas particle and galaxies for the \EAGLE simulation. The authors calculate the neutral hydrogen fraction based on the results of \cite{rahmati2013} who study the column densities of neutral gas in cosmological simulations combined with a radiative transfer calculation. The H$_{2}$ fraction is estimated by using the prescription of \cite{krumholz2013} where the transition between HI and H$_{2}$  depends on the total column density of neutral gas, the gas metallicity, and the interstellar radiation field.
As illustrated in Fig.~\ref{fig:fractionvoids}, star-forming galaxies are  slightly more frequent in void galaxies ($>90$\%) than in exterior regions ($80\%$). Similarly, galaxies with a gas fraction $f_{\rm atom}>0$ are less frequent with increasing void-centric distance, with a galaxy fraction varying from 98 per cent in the inner parts of the void to about 80 per cent in the exterior regions.
When the fraction are estimated for various stellar mass bins, we observe that the decreasing fractions of star-forming galaxies and galaxies with gas as a function of void-centric distance are maintained. Nevertheless, the fractions fall at different rates. For instance the fraction of star-forming galaxies with $M_{*}>10^{10}\Msun$  falls swiftly from $0.92$ at $0.4 r/\Rvoid $ to $0.8$ for $r/\Rvoid >1$, while low and intermediate galaxies ($M_{*}<10^{10}\Msun$) decrease more gradually from $0.96$ at $0.4 r/\Rvoid $   to less than 0.8 for $r/ \Rvoid >1.8$. In the case of the fraction of galaxies with  $f_{\rm atom}>0$, we found that the downward trend is preserved  regardless of stellar mass, with the exception of massive galaxies, for which  the fraction remains nearly constant at $0.92$ for $r/\Rvoid >1$.
Our findings are consistent with the results of \cite{ricciardelli2014}(see their Fig. 8)  which show that SF galaxies decreases with the void-centric distance and for two stellar mass bins ($M_{*}>10^{9.5}\Msun$ and lower than this mass).
The fraction of high mass galaxies in their sample also presents a steeper decreasing trend shown in our massive galaxies.  The authors suggested that galaxies in voids seem to be more efficient in forming stars than galaxies in other regions and highlighting that galaxy evolution in voids is slower with respect to the evolution of the general population. From the theoretical side, \cite{habouzit2020} have shown similar results to our findings with ZOBOV \citep{neyrinck2008} and VIDE \citep{sutter2015} that are void finders relying on a Voronoi Tessellation of the tracer distribution in the \textsc{HorizonAGN}. This simulation has different subgrid physics of galaxy formation as \textsc{EAGLE}.



The higher frequency of star-forming galaxies and galaxies containing hydrogen gas in voids could simply be explained by the bias towards low masses in underdense regions. We will investigate this in the following sections using galaxy subsamples from various environments with the same stellar mass distribution as the inner void sample.

\section{Galaxy properties comparison between subsamples with equal stellar mass distribution}
\label{sec:galprop}

This section examines the effects of stellar mass and environment on galaxy properties. To accomplish so, we define galaxy subsamples from the galaxy parent samples corresponding to different environments, as analysed in the previous section (see Subsection \ref{subsection:galaxysample}).
These subsamples have the same stellar mass distribution as the parent galaxy sample in the inner void regions. We control by stellar mass rather than halo mass since stellar mass is easier to infer from observations. Using subsamples selected by requiring host haloes to match halo mass distribution does not appreciably modify the major conclusions in appendix \ref{app:halosame}.

\begin{figure*}
	\begin{tabular}{cc}
 	\includegraphics[width=1\columnwidth]{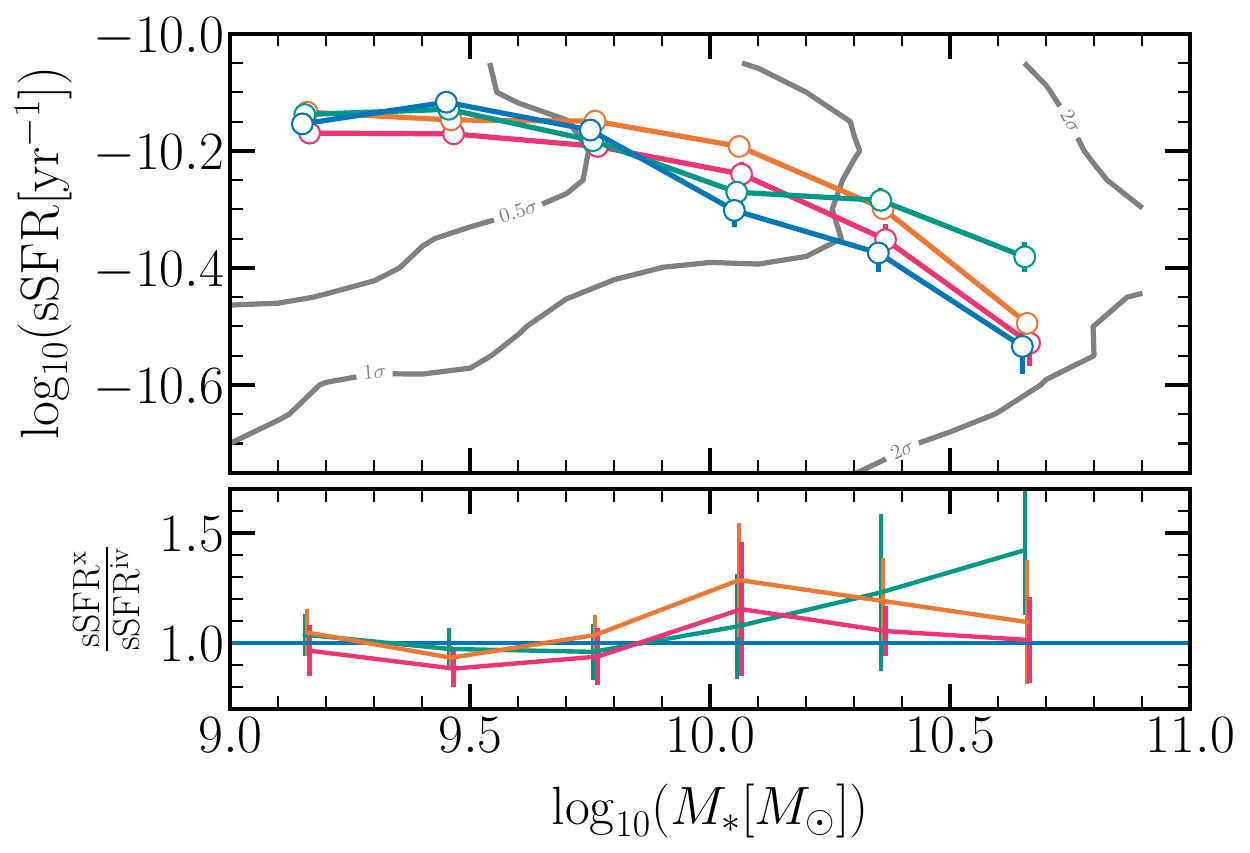} &
	\includegraphics[width=1\columnwidth]{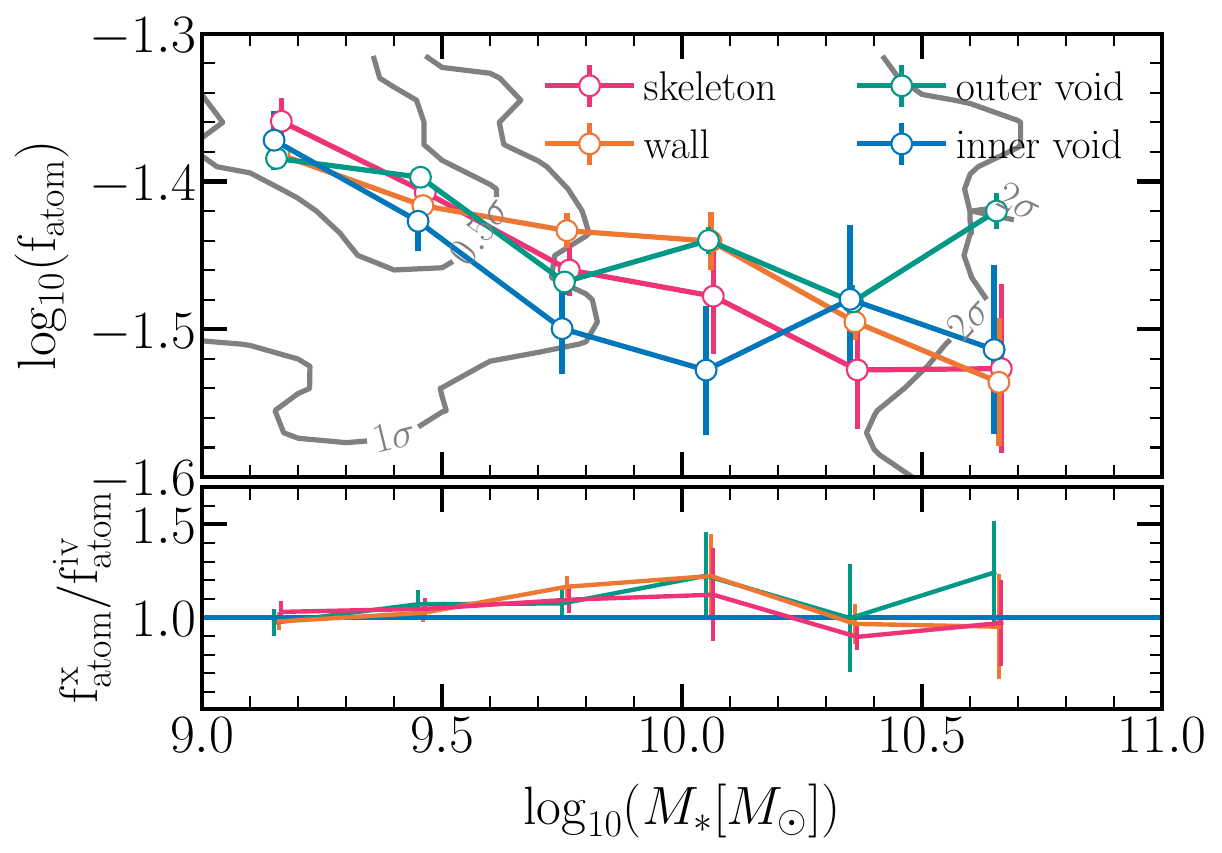}  \\
	\end{tabular}
    \caption{Properties of star-forming galaxies in the subsamples, as the legend indicates. Grey contours represent all the central galaxies in the simulation.  Coloured markers and solid lines represent the mean distribution, and error bars represent jackknife errors using $10$ subsamples. \textit{Left panel:} The mean specific star formation rate (sSFR) as a function of stellar mass.  \textit{Right panel:} The mean Hydrogen gas fraction ($f_{\rm atom}$) as a function of stellar mass, where $f_{\rm atom}= 1.35 M_{\rm HI}/ (M_{*} +1.35 (M_{\rm HI} +M_{\rm H_{2}}))$. The bottom figures show the ratio between the mean sSFR (left panel) and the mean gas atomic fraction (right panel) for each subsample and inner void galaxies.} 
    \label{fig:samestellarmass}
\end{figure*}

\begin{figure}
	\includegraphics[width=1\columnwidth]{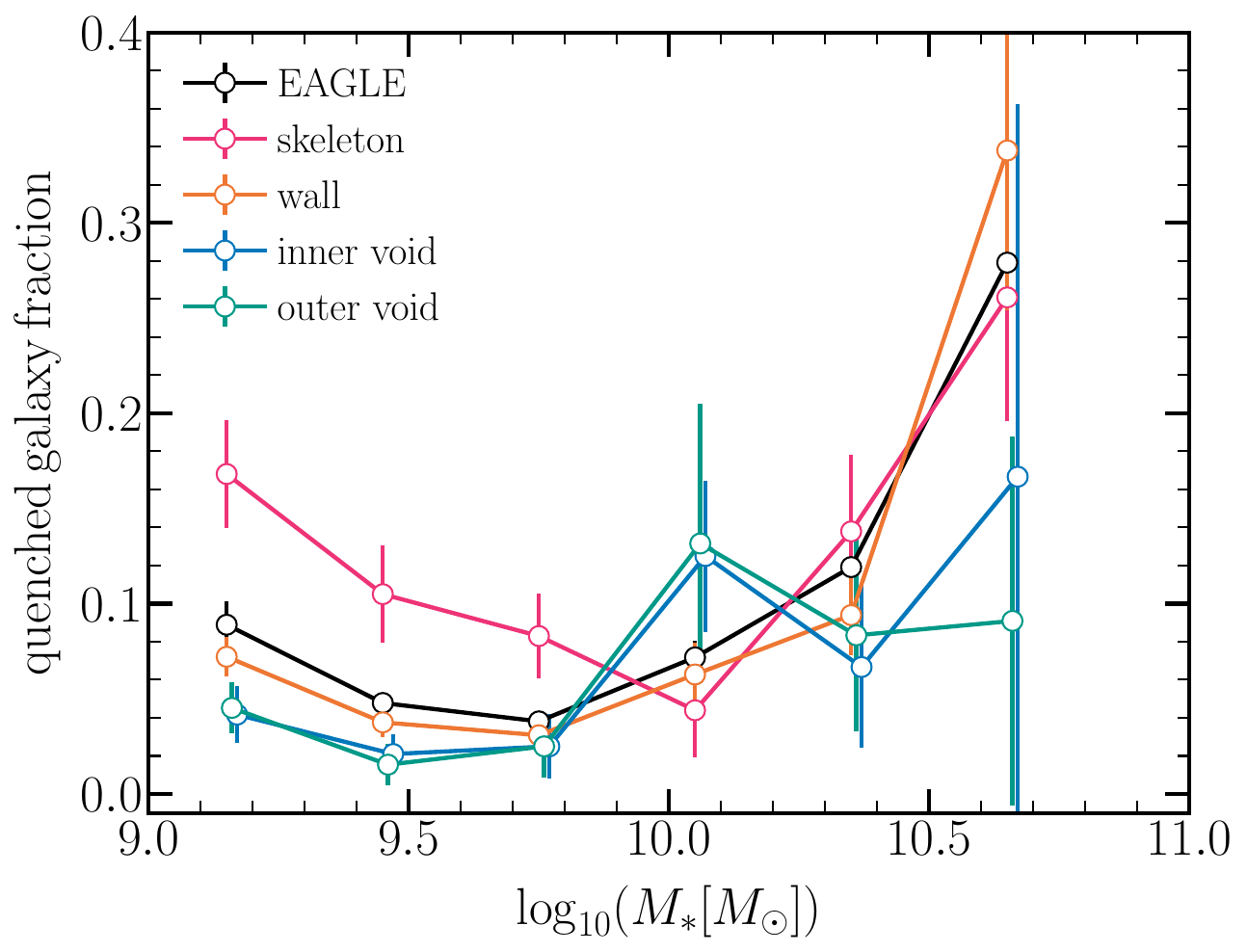}
    \caption{The fraction of galaxies that are quenched (sSFR$<10^{-11.5}\rm yr^{-1}$) as a function of stellar mass for the inner void galaxies and subsamples of denser regions as in the legend is specified. The solid black line represents the quenched galaxy fraction in the \EAGLE simulation. Errorbars correspond to jackknife errors using 10 subsamples. In general, the fraction of quenched galaxies is higher in skeleton and wall galaxies than the inner and outer void galaxies.}

    \label{fig:quenchedfraction}
\end{figure}
\begin{figure}
	\includegraphics[width=1\columnwidth]{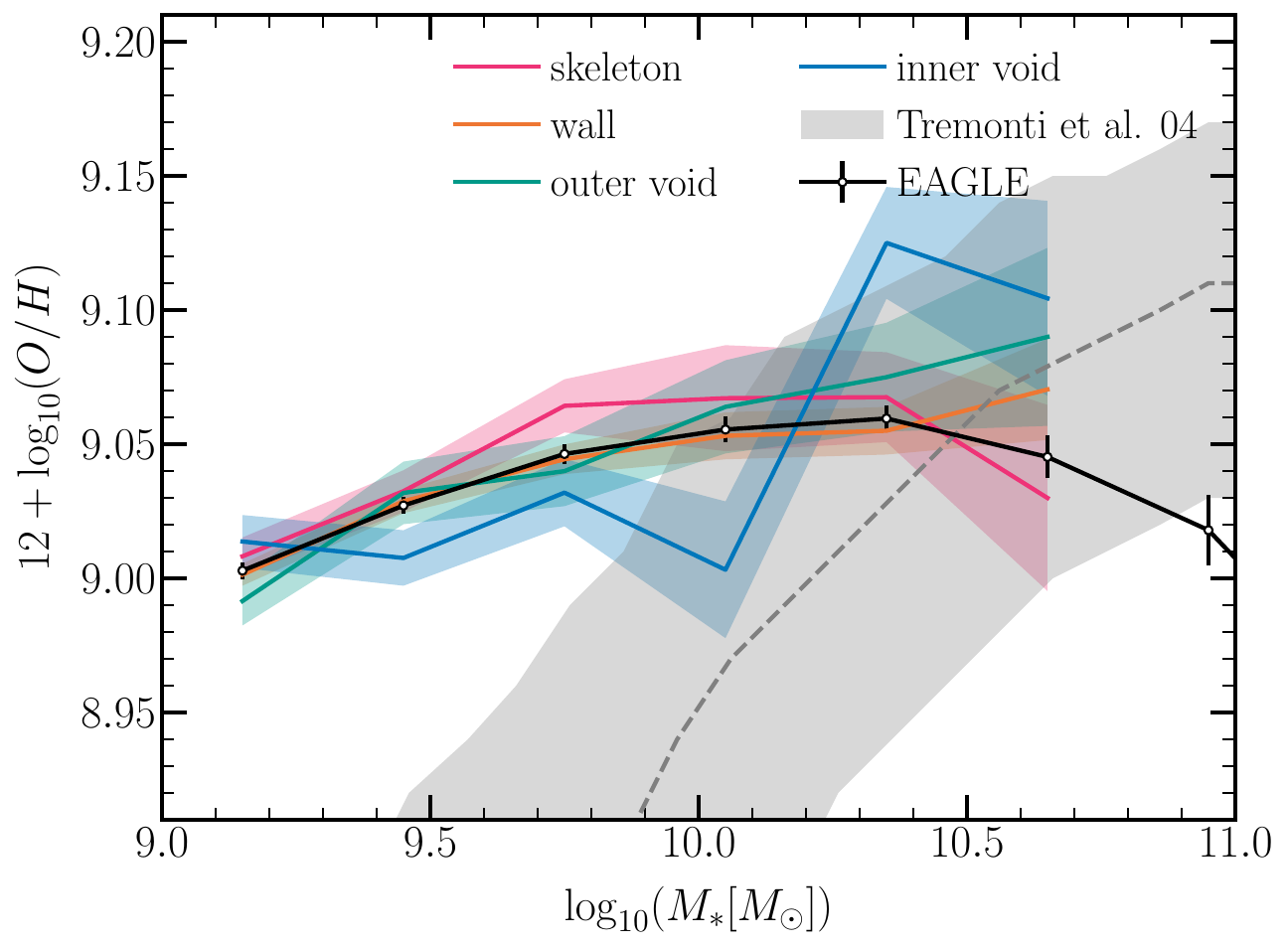}
    \caption{The mean relation between stellar mass and the star-forming gas-phase oxygen abundance.  Shaded regions represent the jackknife errors using 10 subsamples and  the black solid lines  and with circles represent the mean relation with entire simulation.  Grey dotted line and shaded region are observational estimations from \protect\cite{tremonti2004}. Inner outer void galaxies present lower metallicities than skeleton and wall galaxies. The difference is more pronounced at higher mass galaxies.}

    \label{fig:starmass-metallicidad}
\end{figure}
\begin{figure}
	\includegraphics[width=1\columnwidth]{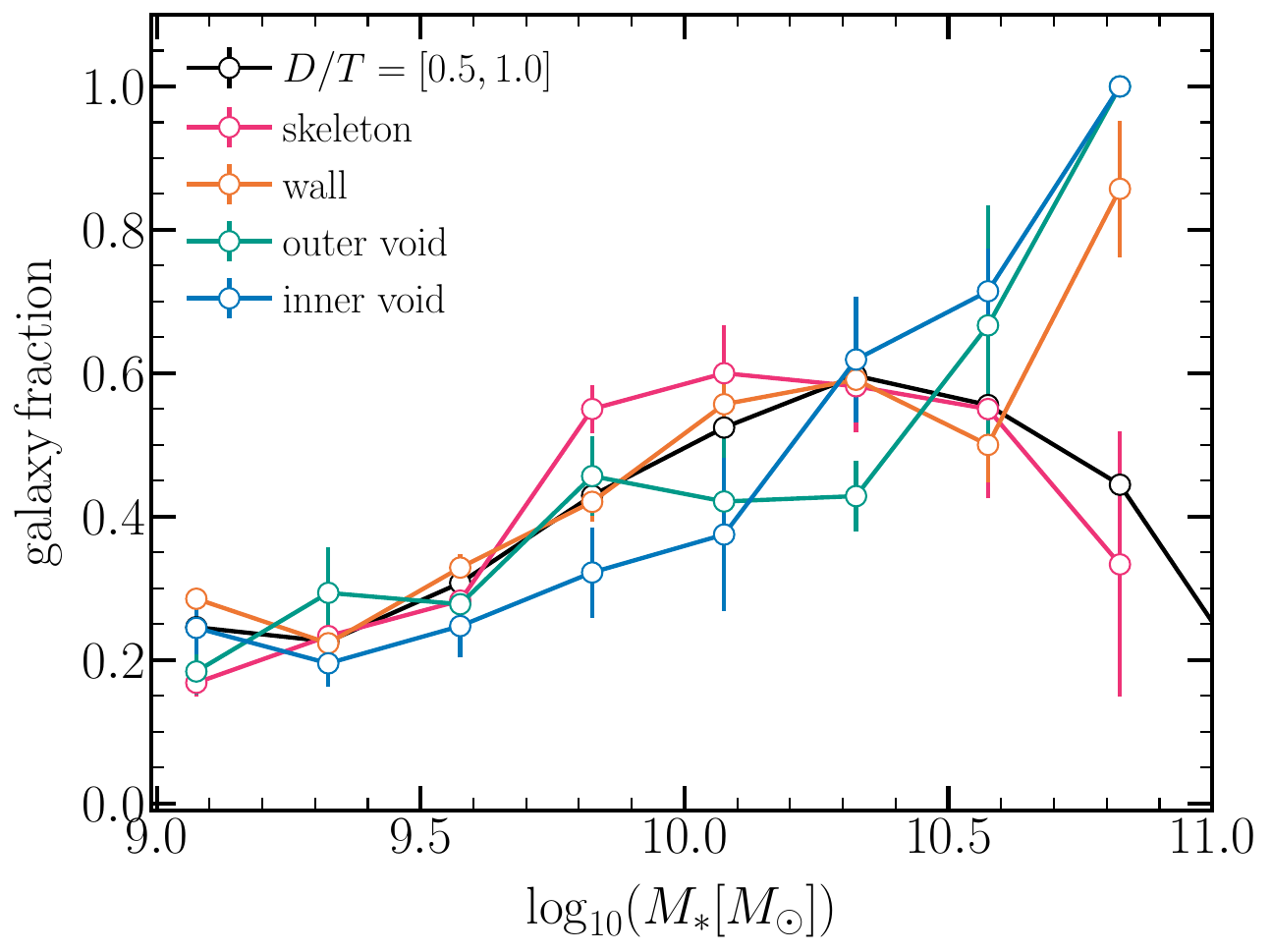} \\
    \caption{The fraction of galaxies that have a dominant stellar disc as a function of stellar mass for the different subsamples.  Errorbars correspond to jackknife errors using 10 subsamples. Inner voids present the highest fraction of discs in galaxies with $M_{*}\geq10^{10.3}\Msun$.}

    \label{fig:morphclass}
\end{figure}

\subsection{Galaxy properties as a function of stellar mass}
\label{subsec:galaxyproperties}
The left panel of Fig.~\ref{fig:samestellarmass} shows the  sSFR-stellar mass plane for each subsample.
The lines with circles and error bars represent the mean distribution and the jackknife errors in the mean estimations of each subsample, respectively. Grey contours denote all the central galaxies in the \EAGLE simulation with a stellar-mass larger than $10^9\Msun$ and sSFR$=10^{[-10,-11.5]}\rm yr^{-1}$. The figure shows that galaxies follow the same sequence regardless of their cosmic web location: low stellar mass galaxies are active ($M_{*}\leq10^{10}\Msun$) near the main sequence of star-forming galaxies, whereas massive galaxies have less star-formation activity, with the exception of massive outer void galaxies ($M_{*}\geq 10^{10}\Msun$) that appear to have higher star-formation activity.
For low mass galaxies  ($M_{*}= 10^{9-9.75}\Msun$) inner void and outer void galaxies are more active than their wall and skeleton counterparts. For intermediate mass galaxies ($M_{*}=10^{9.75-10.25}\Msun$), we find opposite behaviour depending if they inhabit the inner/outer void regions or the other regions: inner and outer void galaxies are slightly less SF active than skeleton and wall galaxies. For massive galaxies ($M_{*}>10^{10.25}\Msun$), there is no discernible variation in the star formation activity with the exception of the outer void galaxies, which have a higher star-formation activity. Note that these differences are modest as depicted in the bottom plot of  the left panel of Fig.~\ref{fig:samestellarmass}, which shows the ratio between the sSFR of each subsample and the inner void sample as a function of the stellar mass and their corresponding jackknife errors. For low stellar mass galaxies, this ratio in SF activity is small (by up to 1 per cent) whereas for intermediate galaxies are up to 10 per cent. For the most massive galaxies, the difference in the SF activity  becomes significant for the outer void galaxies in comparison with the other environments. This difference in the outer void regions seems to be related to the fact that a high fraction of the outer void galaxies reside in small voids ($\Rvoid<10$ pMpc) as we will discuss in Section \ref{sec:discussion}, these voids could be more affected by hot, metal-rich gas due to feedback process.

This is in agreement with \cite{ricciardelli2014}, who using cosmic voids
identified in SSDS DR7, found no significant differences in galaxies in voids and the field that lie in the star-forming main sequence with the same stellar mass (see their Figure 7). However, the authors
found discrepancies in the proportion of star-forming and quenched
galaxies, suggesting that galaxies in dense regions have undergone
fast quenching mechanisms as a result of environmental effects such
as galaxy mergers, which rapidly enhance their star-formation efficiency.

Indeed, as illustrated in Fig.~\ref{fig:quenchedfraction}, the fraction of quenched galaxies (sSFR$<10^{-11.5} \rm yr^{-1}$) varies  for each subsample.  The figure clearly shows that the fraction of quenched galaxies in inner void regions is lower than the fraction of quenched galaxies in subsamples of denser environments for the majority of the stellar mass bins, with the exception of galaxies with $M_{*}=10^{10}\Msun$. For example, a fraction of quenched galaxies of 0.17 is observed in low-stellar mass galaxies ($M_{*}=10^{9.3}\Msun$) in the skeleton regions, whereas only a fraction of 0.04 of low-stellar mass galaxies in the inner and outer void regions are quenched. It is worth noting that for stellar masses $M_{*}\geq10^{10.25}\Msun$, while the inner void galaxy sample contains fewer quenched galaxies than the rest of the subsamples, the difference between galaxies in voids and those in denser environments is more modest. Besides, the fraction of quenched galaxies increases with stellar mass in denser environments, possibly due to quenching mechanisms affecting mostly massive galaxies, such as major mergers and AGN (Active Galaxy Nuclei) feedback. In the case of the wall and particularly, the skeleton regions, there is also a clear turnover with an increasing fraction of quenched galaxies for low-stellar mass galaxies. This could also be due to the action of ram-pressure stripping due to the large-scale gas, which deprives small galaxies of gas \citep{benitez-llambay2013,marasco2016}. It is interesting to note that the small peak of quenched galaxy fraction seen at intermediate stellar masses ($M_{*}=10^{9.75-10.25}\Msun$) in inner and outer voids may be a result of the AGN feedback that becomes effective at this $M_{*}$ range. \cite{paillas2017} demonstrate that feedback processes  contaminated voids with hot, metal rich gas, specially voids with $\Rvoid<10$ pMpc, This is compatible with our findings where above 70 per cent of quenched galaxies at intermediate stellar masses in inner voids are located in small voids ($\Rvoid<10$ pMpc). Also, this is consistent with the low SF activity seen in inner voids at this stellar mass range.

Overall, we found that the fraction of quenched galaxies increases as the void-centric distance increases from $0.04$ in the inner void regions to $0.12$ in the skeleton regions.These trends might be caused by the availability of gas supply to feed star formation, as well as environmental effects. For instance, galaxies living on the outskirts of a void have intermediate local densities, such that the effects of the environment that can quench galaxies are expected to be less important than those located in the skeleton and walls. Galaxies living in the inner parts of a cosmic void, however, might have a more steady evolution dominated by secular evolution with lower star-formation efficiency.
The right panel of  Fig.~\ref{fig:samestellarmass} shows the average atomic gas fraction,$f_{\rm atom}$, for each subsample and inner void galaxies. To emphasise the differences in the HI gas fraction between each subsample and the inner void galaxies, the bottom panel
compares the HI gas fractions of each region to the HI gas fractions of inner void galaxies.  As can be seen, inner void galaxies with $M_{*}\geq 10^{9-9.75}\Msun$ have slightly lower HI gas fractions on average than those in denser regions, although the statistical uncertainties are large. For intermediate stellar masses  ($M_{*}=10^{9.75-10.25}\Msun$),  it is evident that inner void galaxies have the smallest fractions of HI gas, whereas wall galaxies have the largest fractions compared to other places.  Notably, when compared to observations, this is compatible  with a recent study by \cite{dominguez2022} that compared the molecular and atomic gas of control samples of galaxies in filaments and voids using the Void Galaxy Survey (VGS; \citealt{beygu2016}) and HI data from  \cite{kreckel2012} combined with measurements of CO emission lines. The authors found no significant differences across the samples. However, for these intermediate stellar masses, the atomic gas mass fraction is lower in void galaxies than those in filaments.

It is worth mentioning that the highest HI gas fractions seem to be in wall galaxies with $M_{*}=10^{9.75,10.25}\Msun$ in our sample. This is qualitatively consistent with the results of  \cite{janowiecki2017} who, using deep HI observations from the extended GALEX Arecibo SDSS survey (xGASS), found that galaxies with $M_{*}=10^{10.2}\Msun$ residing in groups have higher HI gas fractions than those in isolation.  

\begin{figure*}
    \begin{tabular}{cc}
	\includegraphics[width=1\columnwidth]{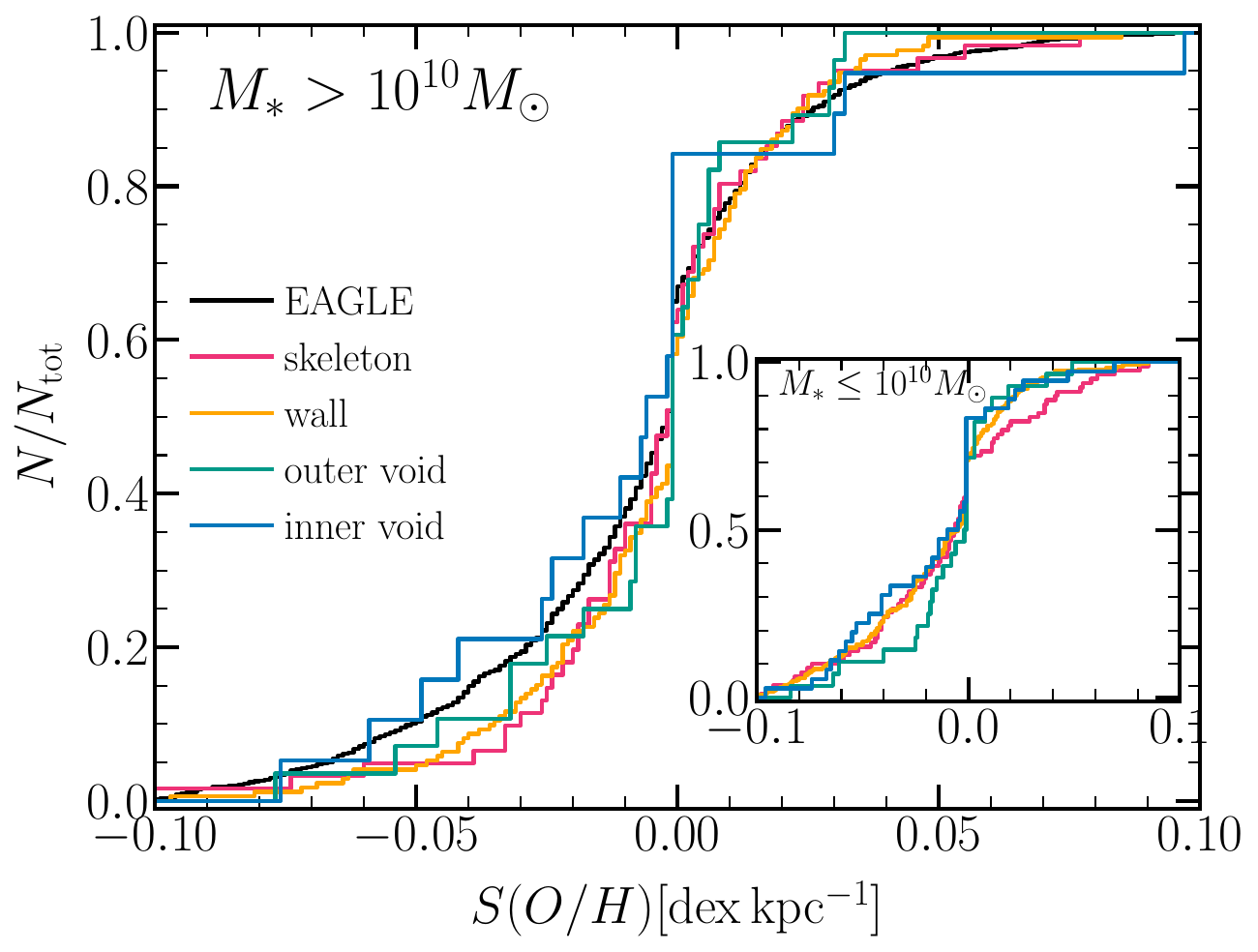}&
	\includegraphics[width=1\columnwidth]{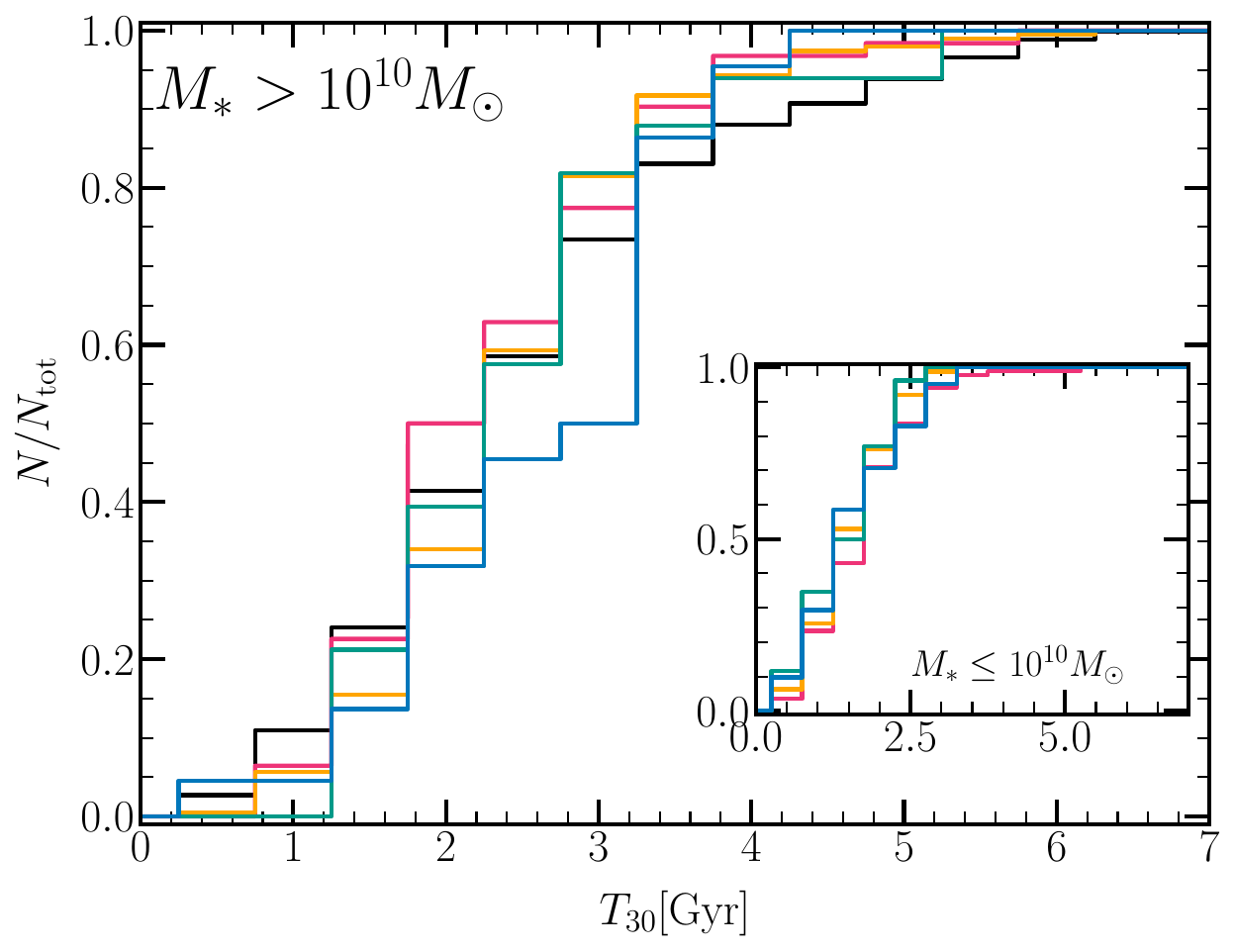} \\
	\end{tabular}
    \caption{\textit{Left panel:} The cumulative distribution of metallicities gradients taking as a proxy of metallicity, star forming gas-phase oxygen abundance ($S(O/H)$) for galaxies  with stellar masses $>10^{10}\Msun$ and in the inset plot for those in stellar masses $\leq 10^{10}\Msun$ for the different subsamples as indicated in the legend. \textit{Right panel:} The cumulative distribution of the age of the youngest 30 per cent of the stellar population ($T_{30}$). Galaxies in inner void galaxies have a higher fraction of negative metallicities gradients than those in other regions.}    \label{fig:discs}
    \end{figure*}

An important property that characterises a galaxy is its average gas metallicity since it encapsulates the assembly history of a galaxy.  In particular, it constrains how the gas has been reprocessed by stars and by different types of exchanges with their surroundings.  Fig.~\ref{fig:starmass-metallicidad} shows the mean relation between the star-forming gas-phase oxygen abundance and the stellar mass for the star-forming galaxies for the subsamples and inner void galaxies. This is the well-known mass-gas phase metallicity relation (MZR). For small galaxies within $M_{*}=10^{[9.0,9.5]}\Msun$, inner void galaxies tend to have higher gas-phase metallicities than those residing in denser regions. This might be consistent with the higher gas accretion rate expected for galaxies in denser regions and the fact that gas accretion tends to bring lower metallicity gas \citep{collacchioni2020,wright2021}. Although, this tendency is weak. For inner void galaxies with $M_{*}\sim 10^{[9.5,10]}\Msun$, they have lower gas-phase metallicities than those galaxies residing in denser regions. In particular, the highest difference appears when inner voids are compared with skeleton galaxies. This is in agreement with the differences found by \cite{pustilnik2011} in dwarf galaxies in the Lynx-Cancer Void using spectroscopic and SDSS data. The authors showed that these galaxies have lower gas-phase metallicities than dwarf galaxies in denser environments. At higher stellar masses ($M_{*}>10^{10}\Msun$), in contrast, the gas-phase metallicity for the inner/outer void galaxies is higher and increases with stellar mass. The gas-phase metallicity for skeleton and wall galaxies flattens with a stellar mass similar to the relation found for the entire central galaxy population.

To understand the shape of the MZR, \cite{deRossi2017} have studied the evolution of the MZR in \EAGLE  for different efficiency of AGN feedback and found that the flattening of the MZR in higher stellar mass is due to the action of strong AGN feedback, which possibly generates metal-rich mass-loaded winds. Comparing the MZR from the simulation to the observational relation from \cite{tremonti2004}, shown by the dashed grey band, we do not find an agreement with observational data. This is a well-known fact and is driven by numerical resolution \citep{schaye2015}. However, the highest resolution simulation (but smaller volume) from the \EAGLE suite has a good agreement with the data (see \citealt{deRossi2017}). We include the median relation of \EAGLE galaxies (solid black line with error bars) as a reference to assess the different levels of enrichment reached by galaxies at different locations.



Finally, we explore the morphology of our subsamples.  The fraction of disc galaxies as a function of the stellar mass is shown in Fig.~\ref{fig:morphclass}. Disc galaxies are defined as those with $D/T \geq 0.5$ where $D/T$ denotes the fractional stellar mass in the disc component, as described in section~\ref{subsec:morphclass}. The figure reveals that the fraction of disc galaxies increases with stellar masses for inner and outer voids, whereas the fraction of disc galaxies for skeleton and wall galaxies, peaks at $M_{*}\geq 10^{10}\Msun$, following the trend of all centrals in the \EAGLE simulation (the grey line). For low-mass galaxies ($M_{*}<10^{9.75}\Msun$), the proportion of disc galaxies does not vary much between subsamples. However, for galaxies with an intermediate stellar mass  ($M_{*}=10^{9.75-10.25}\Msun$),
the difference in the disc galaxy fraction is higher, with inner voids presenting the lowest fraction of discs. In contrast, massive galaxies ($M_{*}\geq10^{10.25}\Msun$),  in inner and outer voids exhibit an overabundance of discs compared to those in other locations, with disc fractions exceeding 60 per cent.   The highest fraction of massive disc-dominated galaxies are in agreement with the observational results of \cite{rojas2004} who report a higher frequency of disc-like galaxies in voids using SDSS.  If void galaxies were also isolated in the past, the disc-like morphology would be expected in voids. By examining the merger histories of galaxies, it may be feasible to comprehend the origin of the smallest proportion of disc galaxies with lower stellar mass \citep{lagos2018,rosito2019,lagos2022}. In Section \ref{sec:assembly}, we analyse the merger histories of the subsamples, where no significant differences are detected across regions. To understand the lowest disc fraction,  however, it would be necessary to study  how the mergers experienced by galaxies  shape  the morphology  across time, which is out of the scope of this paper.
Overall, however, there is no large disparity in the fractions of disc galaxies between the inner void regions (29 per cent are disc galaxies) and subsamples from denser regions (30 per cent are disc galaxies), but the dependence of the frequency of disc galaxies at a given stellar mass is clearly different. Our findings of massive galaxies are in agreement with the observational results of \cite{rojas2004} who report a higher frequency of disc-like galaxies in voids using SDSS.

\subsection{The halo mass-stellar mass relation in subsamples}
\begin{figure}
	\begin{tabular}{c}
	\includegraphics[width=1\columnwidth]{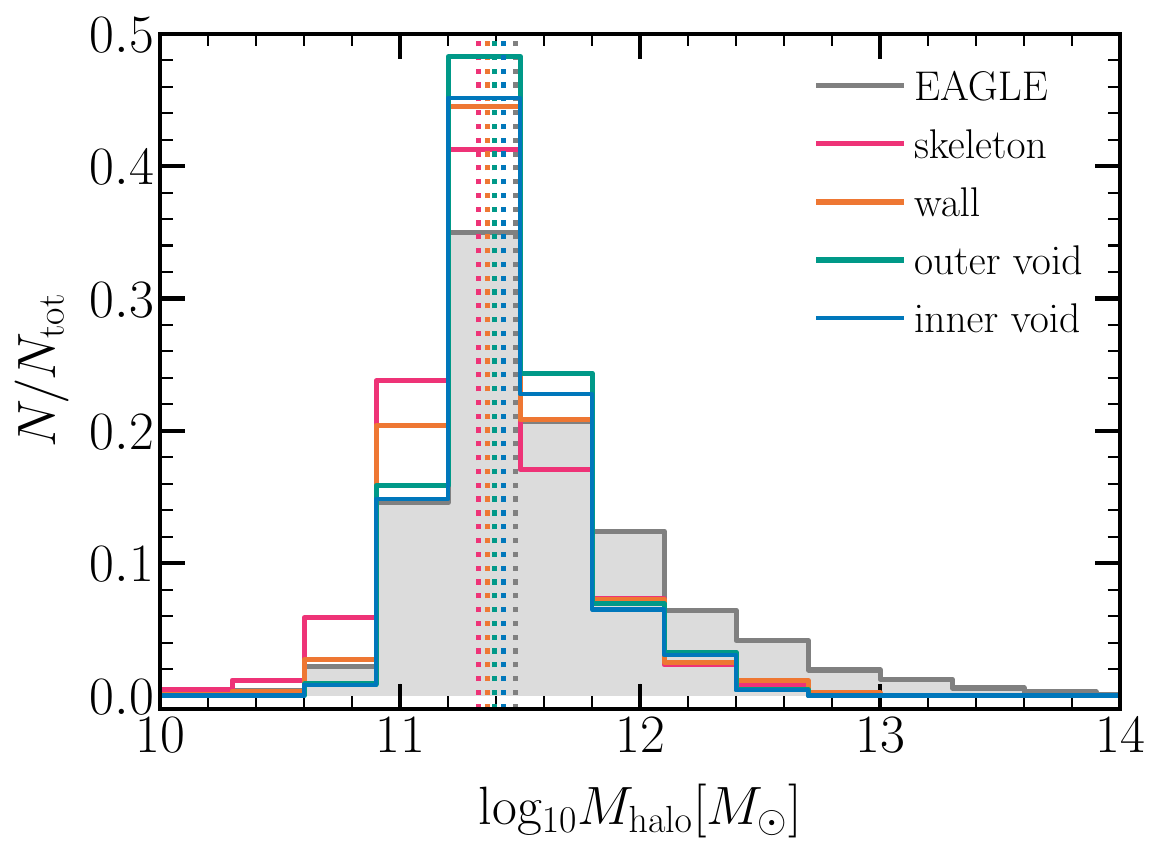}\\
	\includegraphics[width=1\columnwidth]{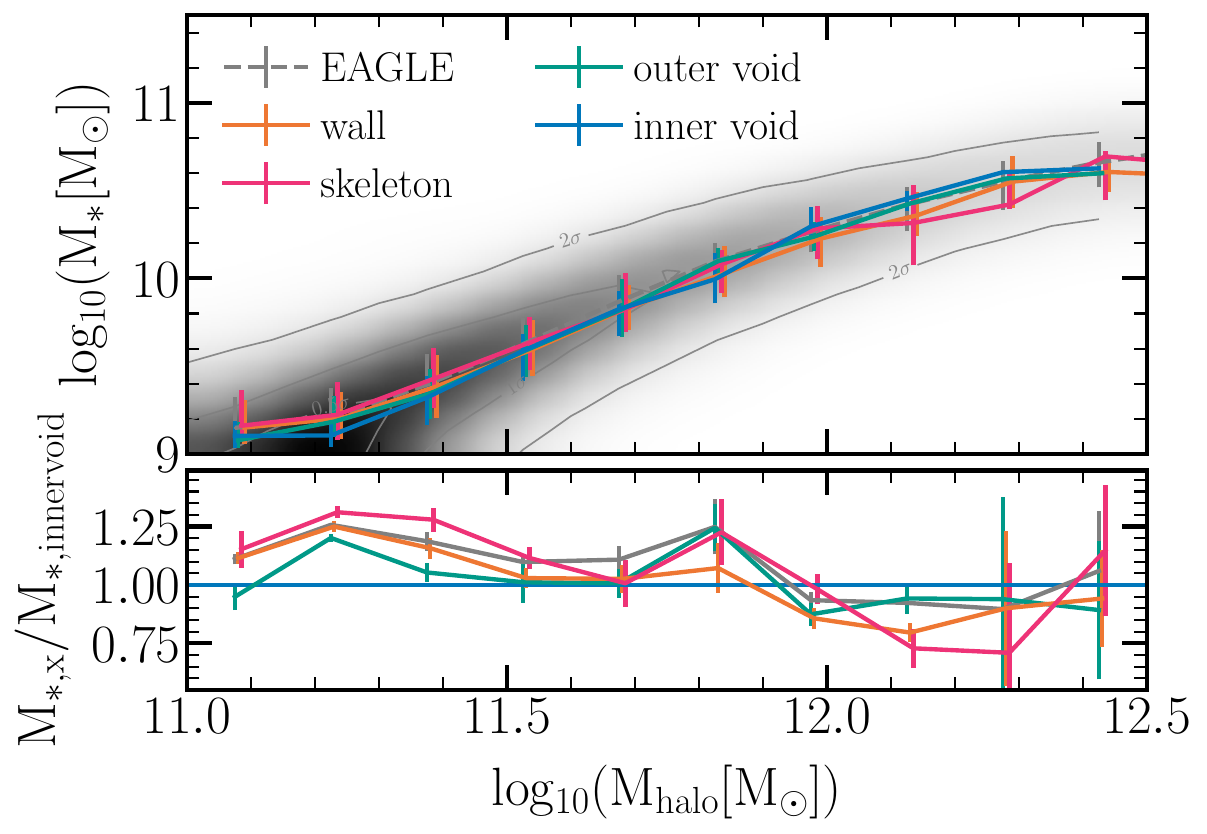} \\
	\end{tabular}
    \caption{ The halo mass (upper panel) distributions and the halo mass and stellar mass relation (bottom panel)  for the subsamples with the same stellar distribution as the one in the inner void sample. Vertical lines represent the median of each distribution. Grey colour and the diffused density map and contours represent the distribution of all central galaxies in the simulation. Error bars represent the $20^{\rm th}$ and $80^{\rm th}$ percentiles of each sample. The ratio between the stellar mass of each subsample and the median distribution of inner void galaxies for a given halo mass are shown in the bottom figure with errorbars corresponding to jackknife errors using 10 subsamples. Inner void haloes hosting lower stellar mass galaxies in haloes with $M_{\rm halo}10^{[11-12.0]}\Msun$ are preserved in the subsamples.}
    \label{fig:gmf_samestellarmass}
\end{figure}

To understand the differences found in the average metallicities and fraction of hydrogen gas, we explore the halo mass-stellar mass relation.
The halo mass distribution in the upper panel of Fig.~\ref{fig:gmf_samestellarmass} shows that  inner voids haloes are more massive than those in other environments. For instance, skeleton galaxies and wall galaxies have wider halo mass distributions with a slightly smaller median halo mass (log$_{10}(M_{\rm halo}[\Msun])=11.3\pm 0.2$ and $11.4^{+0.1}_{-0.2}$  with the $25^{\rm th}-75^{\rm th}$ percentiles respectively), whereas inner and outer voids present slightly more massive haloes (log$_{10}(M_{\rm halo}[\Msun])=11.4\pm 0.2$ and $11.4^{+0.2}_{-0.1}$ respectively). A KS test between the halo mass distributions of each subsample and the one of inner void galaxies yields p-values smaller than $0.15$. This suggests that controlling stellar mass creates various halo populations in different environments.  We note that by enforcing the same stellar mass distribution, we are missing massive haloes that are present in the \EAGLE simulation (see the grey solid histogram in the middle panel), where the halo mass distribution peaks at $10^{11.5}\Msun$ for the entire halo population. Note, however, that similar halo masses do not necessarily imply equal formation histories as it has been shown that inside voids, galaxies (and haloes) are significantly more biased with respect to the mass than in the field \citep{pollina2019}.

The bottom panel of Fig.~\ref{fig:gmf_samestellarmass} shows the halo mass-stellar mass relation. The panel shows that the trend found in the parent galaxy samples is preserved (see Fig.\ref{fig:gmf}): haloes within voids have a lower stellar mass content than their analogues in denser environments at a fixed halo mass between $10^{11}\Msun$ and $10^{12}\Msun$. This is demonstrated in the bottom subpanel of Fig.~\ref{fig:gmf_samestellarmass}, which compares the median stellar mass of the galaxy subsamples to the median stellar mass in void galaxies (blue line) for a given halo mass. The plot indicates that ratios $>1$ appear, with the largest ratio approaching $1.3$ in haloes of $M_{\rm halo} =10^{11.25}\Msun$ hosting skeleton galaxies. We observe the opposite trend for the most massive haloes ($M_{\rm halo}>10^{12}\Msun$), although there are a few of these haloes in voids. We observed that this flip in the halo mass-stellar mass relation has been observed before in some galaxy formation models (see  \citealt{artale2018} for EAGLE \& \citealt{zhang2021}) and not found in others (\citealt{zehavi2018} for SAMs and \citealt{artale2018} for the original Illustris), hence it is a question that remains open.

In general,  void galaxies with $M_{\rm halo}<10^{12}\Msun$ have a slightly lower stellar mass than galaxies located in denser environments. Something worth noting is that the subsamples of wall and skeleton galaxies have more low mass haloes as seen in the halo mass distributions in the upper panel of Fig.~\ref{fig:gmf_samestellarmass} and this is caused by the fact that the halo sample is only complete for galaxies with a stellar mass larger than $10^9\Msun$. For that reason, we only show the median halo-stellar mass relation in haloes with masses larger than $10^{11}\Msun$ for all the subsamples.
Our findings are consistent with the assembly bias predictions \citep{zehavi2018,artale2018}, which indicate that massive haloes in less dense environments contain less stellar mass. Furthermore, it is compatible with the findings of \cite{alfaro2020} who, using the \textsc{tng300} simulation \citep{springel2018}, found that haloes in inner voids have less stellar mass content than haloes of the same mass living in denser environments. Note that these works study a wider range of halo masses, whereas our subsamples contain haloes with  $M_{\rm halo}<10^{12.5}\Msun$.

\subsection{Gas-phase metallicity gradients and young stellar population}
\label{sub:gradients}

Chemical elements are spread in-homogeneously throughout galaxies, according to studies of metallicity gradients in galaxies. Metallicity gradients in galaxies are shaped by the mechanisms responsible for gas redistribution, such as torques caused by bars or inflows and outflows induced by SN and AGN feedback \citep{tissera2016,ma2017,hemler2021}. Mergers and interactions can also aid in this process by altering gas properties, mixing chemical elements, and triggering intense star-formation activity \citep{rupke2010,sillero2017}.

In this section, we concentrate on the oxygen abundance gradients of the star-forming gas in the galaxy subsamples with the same stellar mass distribution. The metallicity profiles of galaxies with more than 100 star-forming gas particles in their disc components are estimated. This last condition reduces the number of galaxies with measured gradients to $63$  in the inner void sample, whereas the outer void, wall and skeleton subsamples include $50$, $438$ and $156$ galaxies, respectively.  
\cite{tissera2019,tissera2021} study the metallicity gradients in the \EAGLE simulations. The authors find a large diversity of gas-phase oxygen gradients as seen in observations, with $\sim 40$ per cent of them being positive. Positive gradients in galaxies could be driven by external and internal processes. Interactions with neighbouring galaxies have been identified as an external process, as well as, stellar bars and inflows from AGN and SN feedback which are examples of secular processes that could redistribute the gas in galaxies.
 In particular, \cite{tissera2019} find that galaxies in \EAGLE exhibit a weaker relation between gas-phase oxygen abundances gradients and stellar masses for galaxies that have experienced mergers or strong SN feedback that regulates the star-formation activity than those with quiet merger histories (see also \citealt{tissera2021}).
\cite{tissera2019} also investigates the possible effect of the environment on the gas-phase metallicity gradients,  using the neighbour density (within $500$ kpc) as a proxy for the environment and finding no clear trend.

In this analysis, the classification of the environment is defined differently in terms of the distance to the nearest cosmic void,  allowing us to investigate its impact in more depth.
The left panel of  Fig.~\ref{fig:discs} shows the cumulative distribution of the oxygen abundance gradients of star-forming gas ($S(O/H)$) for the subsamples, which are divided into two stellar mass bins $M_{*}>10^{10}\Msun$ and $M_{*}\leq 10^{10}\Msun$ (inset panel). Inner void galaxies appear to have more negative gas-phase metallicity gradients ($S(O/H)<0$), with $84$ ($80$) per cent of galaxies in the highest (lowest) stellar mass bins. In contrast, galaxy subsamples in denser environments have lower fractions of negative metallicity gradients with $0.51$ ($0.71$ for the lowest stellar mass bin), $0.58$ ($0.72$), $0.61$ ($0.80$) for skeleton, wall, outer void galaxies, respectively, regardless of the stellar mass bin.


The right panel of Fig.~\ref{fig:discs} depicts the cumulative $T_{30}$ that is defined as the lookback time when the last 30 per cent of the stellar population was formed in the disc, after the last merger. The subsamples have been divided into two stellar mass bins:$>10^{10}\Msun$ and $\leq 10^{10}\Msun$ (inset panel).  This parameter may be associated with the input of chemical elements and/or the triggering of outflows, which may have influenced the chemical abundance profiles on the discs in the recent past \citep{tissera2019}. The figure indicates that the youngest stellar populations in massive inner void galaxies (median $T_{30}= 3.0_{-1.25}^{+0.5}$ Gyr) are slightly older than in denser environments (median $T_{30} = 2.25\pm 0.5$ Gyr for the rest of the subsamples). This is compatible with the scenario where secular mechanisms regulate the evolution of massive galaxies in voids. However, as we will see in the next section, not only the youngest population but the entire stellar population is older in inner voids than in massive galaxies ($M_{*}>10^{10}\Msun$) in denser environments.
In low-stellar mass galaxies, there is no significant difference in the age of the youngest stellar population between inner void galaxies (with a median and the $25^{\rm th}-75^{\rm th}$ percentiles of $T_{30}=1.25_{-0.5} ^{+1.0}$ Gyr) and the rest of the galaxy subsamples ($T_{30}= 1.75 \pm 0.5 $ Gyr for outer void and skeleton galaxies and wall galaxies $T_{30}=1.25\pm 0.5$ Gyr). Our findings may be the outcome of the assembly history of galaxies, which we will study in the next section.

\section{The assembly history of galaxies}

\label{sec:assembly}


\subsection{Merger histories}

\begin{figure*}
	\begin{tabular}{cc}

	\includegraphics[width=1\columnwidth]{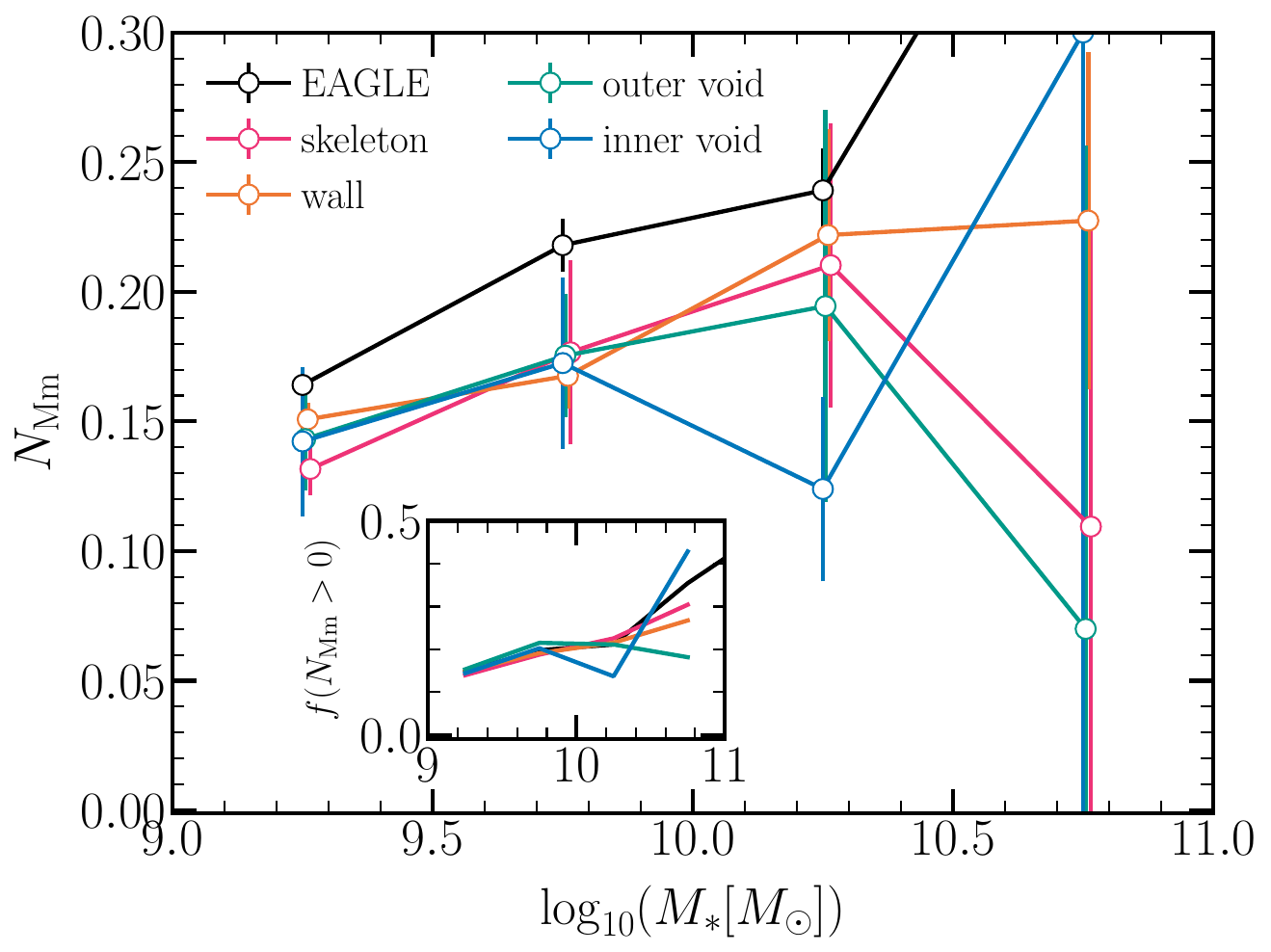}
	&
	\includegraphics[width=1\columnwidth]{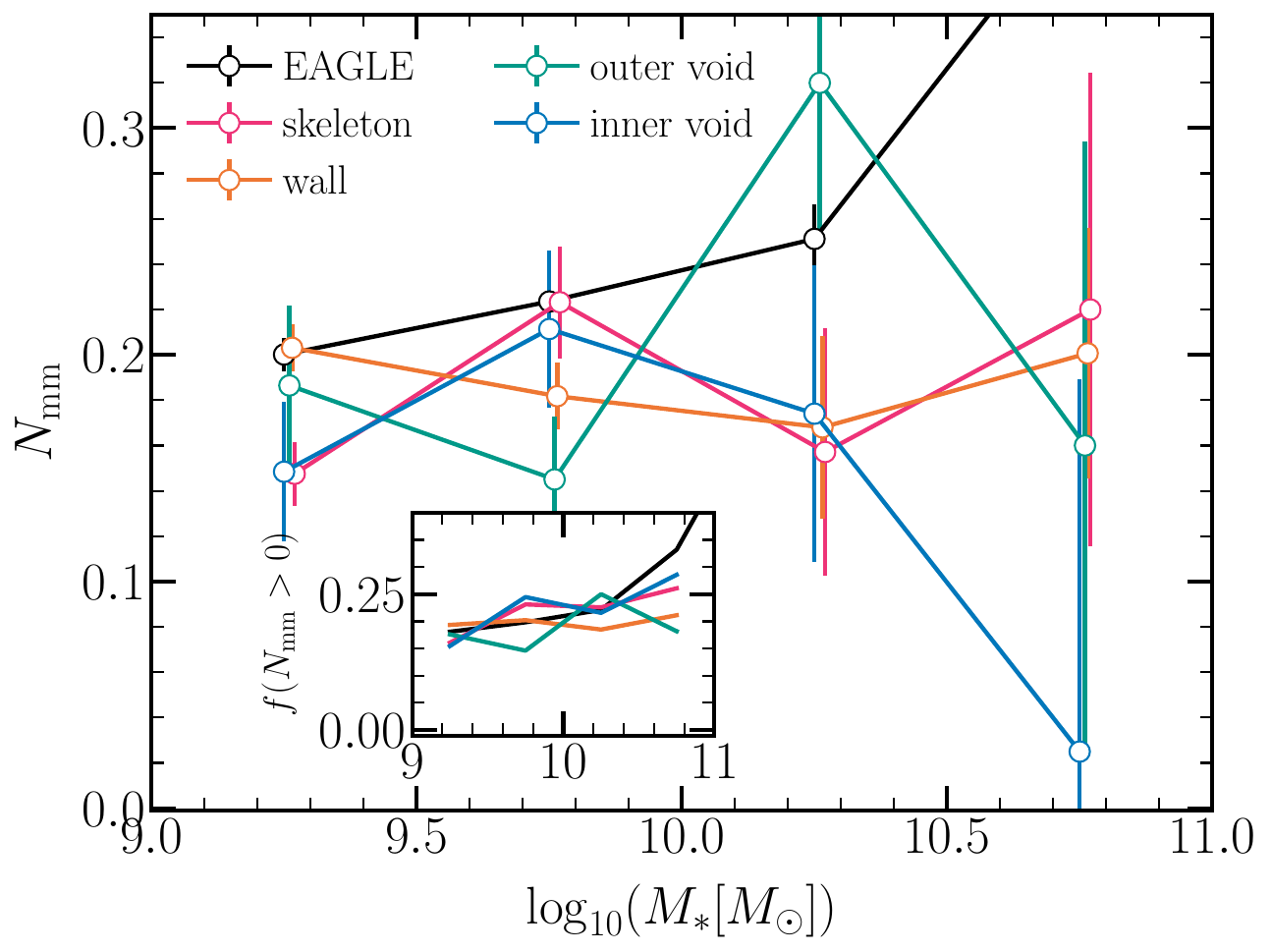} \\

	\end{tabular}
    \caption{ The mean  number of major (left panel) and minor mergers (right panel) as a function of stellar mass for the different environments. Errorbars correspond to jack knife errors using 10 subsamples. The insets show the fraction of galaxies that experienced at least one major/minor merger as a function of stellar mass.}
    \label{fig:mergersmstar}
\end{figure*}

\begin{figure*}
	\begin{tabular}{c}

	\includegraphics[width=2\columnwidth]{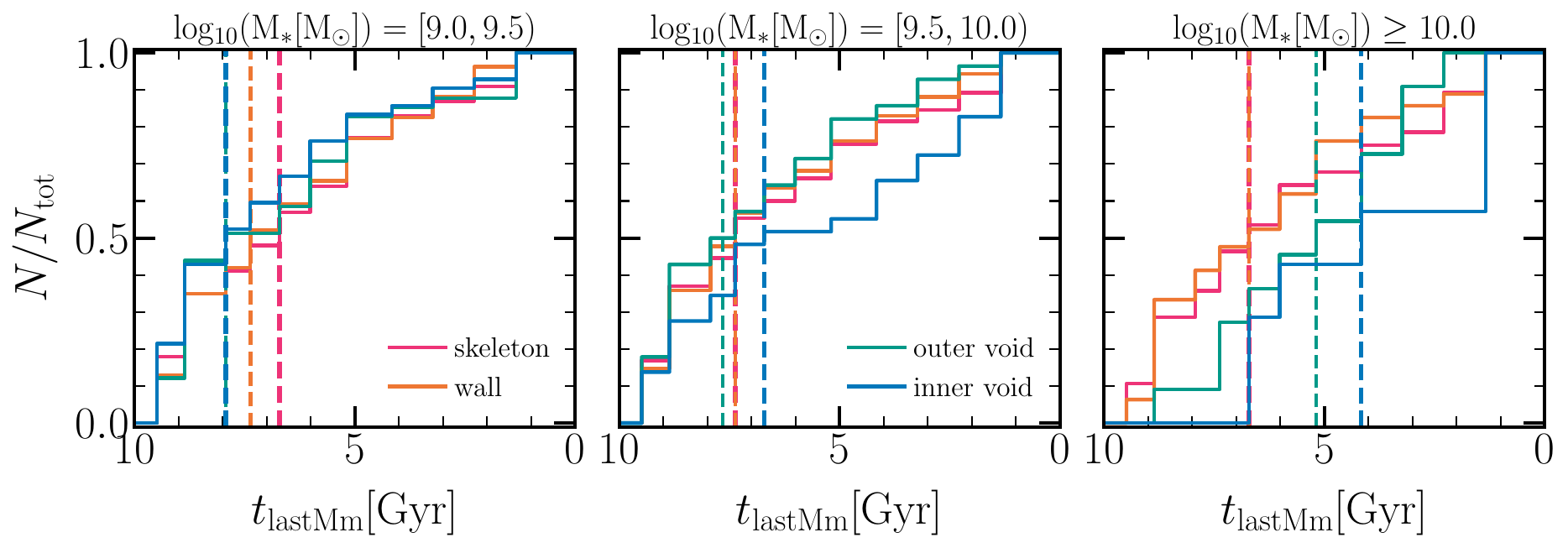}
	\\
	\includegraphics[width=2\columnwidth]{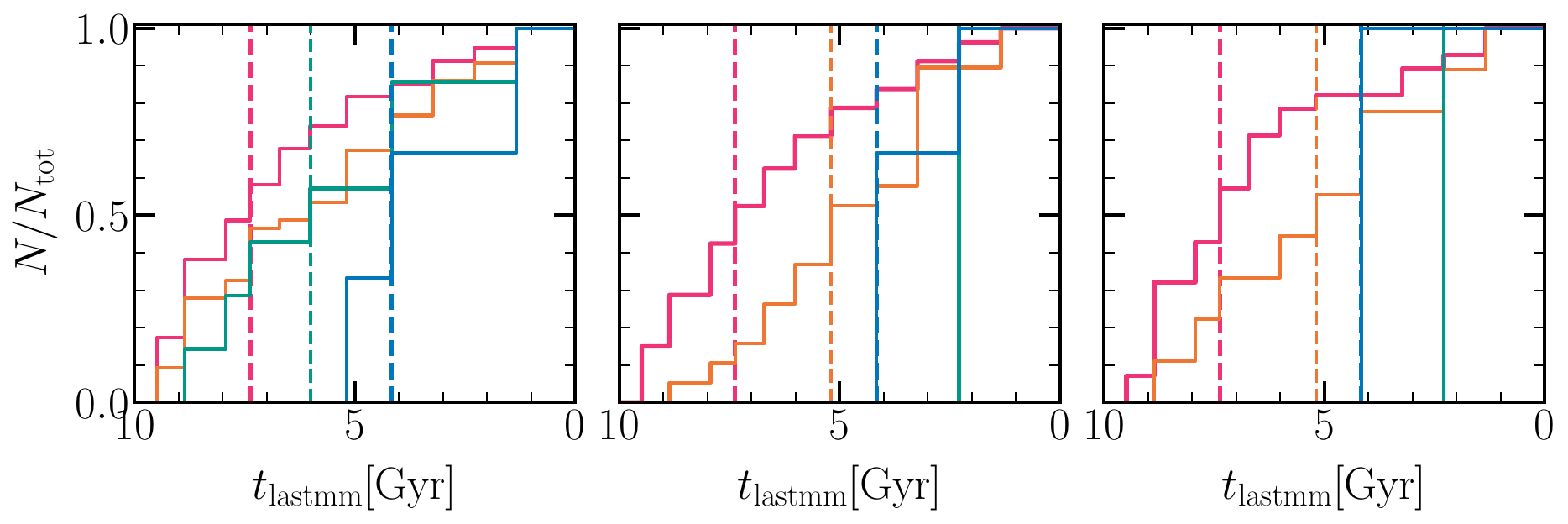} \\

	\end{tabular}
    \caption{ The cumulative distribution of the lookback time when a galaxy underwent its last major (top panels)/minor merger (bottom panels) for the different subsamples and three stellar mass bins, as the figure specifies. The fraction of galaxies that experienced a major (minor) merger oscillates between $\sim10$ ($20$) and $\sim50$ ($25$) per cent for a given stellar mass (see Fig. \ref{fig:mergersmstar}). Dashed lines represent the median of each distribution. Massive galaxies  experienced their last major/minor merger later in inner voids than massive galaxies in denser environments. Low massive galaxies in inner voids experienced their last major merger earlier than galaxies in other environments, but their last minor merger occurred later than galaxies in denser environments}
    \label{fig:lastmergers_ssm}
\end{figure*}

In this section, we compare the merger histories of the subsamples of the outer, wall and skeleton galaxies to those of the inner void galaxies with the same stellar mass distributions. Mergers are classified as major when the stellar mass ratio of the secondary galaxy to the primary galaxy, $\mu$, exceeds $0.25$, whereas for minor mergers $\mu$ takes values between 0.1 and 0.25. To measure $\mu$, we take the ratio of the stellar masses of the galaxies in the last snapshot in which both galaxies are identified as individual structures by  SUBFIND.

We explore the cumulative number of major and minor mergers that galaxies have experienced from $z=1.74$ ($10$ Gyrs ago) to $z=0$  as a function of stellar mass in the different subsamples shown in Fig.~\ref{fig:mergersmstar},  finding that, the number of major and minor mergers increases with increasing stellar mass regardless of the environment. This is consistent with the findings of \cite{lagos2018} who used \EAGLE to determine that the fraction of galaxies that experienced at least one major/minor merger increases with stellar mass. We confirm a similar increasing trend with increasing stellar mass, as shown in the insets of Fig.~\ref{fig:mergersmstar}. Due to the huge spread, there is no noticeable difference between the number of major and minor mergers in different environments. This is expected, given that the subsamples of the different regions match the stellar mass distribution seen in inner void regions. This constrains the merger histories of galaxies in denser regions. In addition, we only consider central galaxies,  whereas taking into account satellite galaxies will reveal differences in the merger histories, as the fraction of satellite galaxies living, for instance, in skeleton galaxies exceeds 50 percent, whereas the fraction of inner void satellite galaxies reaches 4 percent of inner void galaxies.
We note, however, that there are some small differences between the mean number of major/minor mergers of galaxies with $M_{*}=10^{9.5-10.5}\Msun$ in different large-scale environments, although the error are large.   We find no substantial variation in the number of major/minor mergers with regard to the environment for galaxies with $M_{*}\lsim 10^{9.5}\Msun$, partially because they all have a small number of mergers, as expected.
\begin{table}
    \caption{Median lookback time of the latest major merger, $t_{\rm Mm}$[Gyr], and minor merger, $t_{\rm mm}$[Gyr], events for galaxies divided in three different stellar mass intervals in each defined environment. The supra index and under index values represent the difference between  the median and $25^{\rm th}$ and $75^{\rm th}$ percentiles respectively.} 
    \label{tab1}
    \begin{tabular}{ccccc}
    \hline
    &                           &$10^{[9.0,9.5)}\Msun$&$10^{[9.5,10)}\Msun$ & $ \geq 10^{10}\Msun$\\
    \hline
   $t_{\rm Mm}$ Gyr & inner void& $7.9_{+0.9}^{-1.9}$&$6.7_{+2.1}^{-4.4}$&$4.2_{+2.2}^{-2.8}$\\
                    & outer void& $7.9_{+0.9}^{-2.7}$&$7.6_{+1.2}^{-2.5}$&$5.2_{+1.8}^{-1.5}$\\
                    & wall      & $7.4_{+1.5}^{-2.2}$&$7.4_{+1.5}^{-2.2}$&$6.7_{+2.1}^{-1.5}$\\
                    & skeleton  & $6.7_{+2.1}^{-1.5}$&$7.4_{+1.5}^{-2.2}$&$6.7_{+2.2}^{-2.8}$\\
   \hline
   $t_{\rm mm}$ Gyr & inner void& $4.2_{+0.5}^{-1.4}$&$4.2_{+0.0}^{-1.9}$&$4.2_{+0.0}^{-0.0}$\\
                    & outer void& $6.0_{+1.6}^{-1.8}$&$2.3_{+0.0}^{-0.9}$&$2.3_{+0.0}^{-0.0}$\\
                    & wall      & $6.0_{+2.8}^{-1.8}$&$5.2_{+1.2}^{-2.0}$&$5.2_{+2.2}^{-1.0}$\\
                    & skeleton  & $7.4_{+1.5}^{-1.4}$&$7.4_{+1.5}^{-1.4}$&$7.4_{+1.5}^{-1.4}$\\
   \hline
    \end{tabular}
\end{table}

Although the cumulative number of mergers does not vary strongly in galaxies of different environments, the lookback time of the major/minor mergers does. The top panels of Fig.~\ref{fig:lastmergers_ssm} show the cumulative lookback time distribution of the last major mergers in the last $10$ Gyrs for each subsample. We classified galaxies  into three stellar-mass bins: $10^{[9,9.5)}\Msun$, $10^{[9.5,10)}\Msun$ and  $\geq 10^{10}\Msun$. Table~\ref{tab1} also shows the median lookback time of the last major mergers and the $25^{\rm th}$ and $75^{\rm th}$ percentiles of the distributions. Inner void galaxies with $M_{*}\geq10^{9.5}\Msun$ have experienced their last major merger later than galaxies in denser environments. However, the trend is reversed for galaxies with  $M_{*}<10^{9.5}\Msun$: inner void galaxies have experienced their last major merger earlier than the rest of the galaxies.
We should also point out that the distribution of $t_{\rm Mm}$ for massive galaxies($M_{*}\geq 10^{10}\Msun$) overall has lower values than galaxies with lower stellar mass regardless of the environment (see the median in Table~\ref{tab1}), consistent with hierarchical growth \citep{delucia2007}. In particular, massive galaxies in inner voids experienced their last major merger later than massive galaxies living in denser environments.

Variations in the lookback time distributions of minor mergers are also seen as a function of the environment. The bottom panels of Fig. ~\ref{fig:lastmergers_ssm} illustrate the cumulative lookback time distribution for the last minor merger. We see that the median lookback time of the last minor merger in inner and outer voids galaxies occurred later than in skeleton and wall galaxies (see Table~\ref{tab1}).
When we combine all of our findings, we discover differences in the merger histories of galaxies in different environments.

\subsection{Evolution of galaxy properties}
\begin{figure*}

	\includegraphics[width=2\columnwidth]{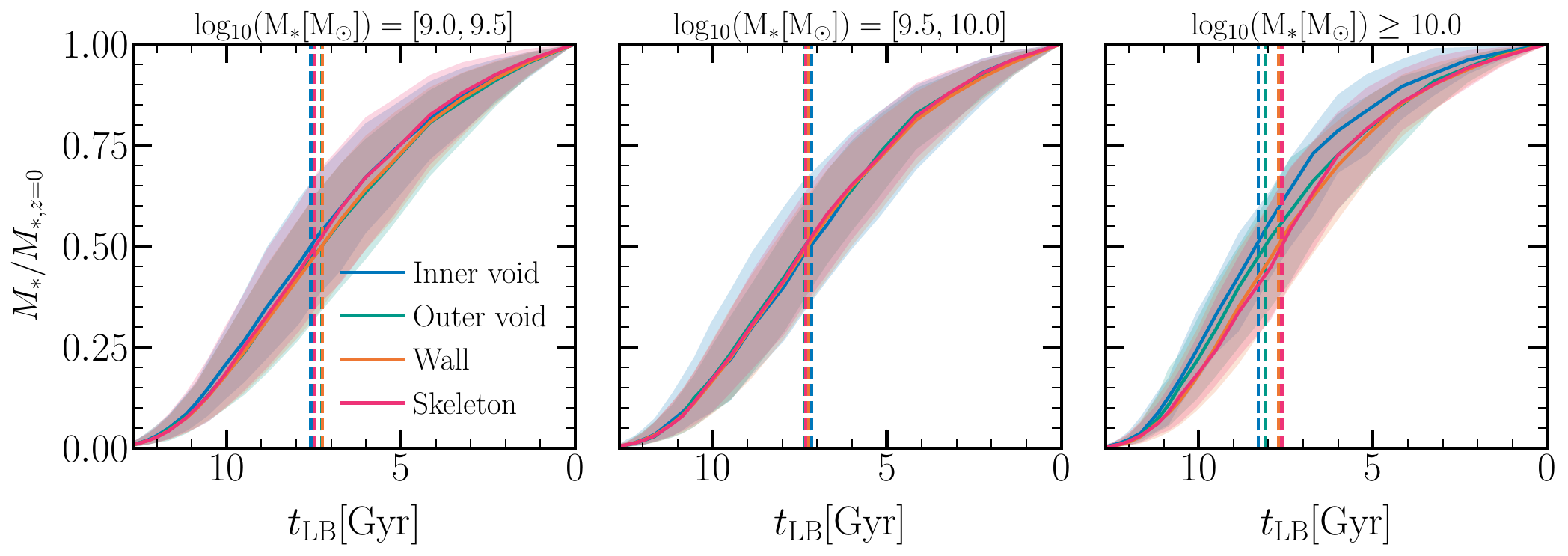} \\

    \caption{The stellar mass relative to the final stellar mass at $z=0$ as a function of the lookback time. Dashed lines represent the formation time of the galaxy, defined as the lookback time when  50 per cent of the stellar mass at $z=0$ was assembled. The assembly history of lowest and highest mass galaxies  seems to form earlier in inner void galaxies than in denser environments, apart from galaxies of intermediate stellar mass in which there is no significant difference.}    \label{fig:galaxyassembly}

    \includegraphics[width=2\columnwidth]{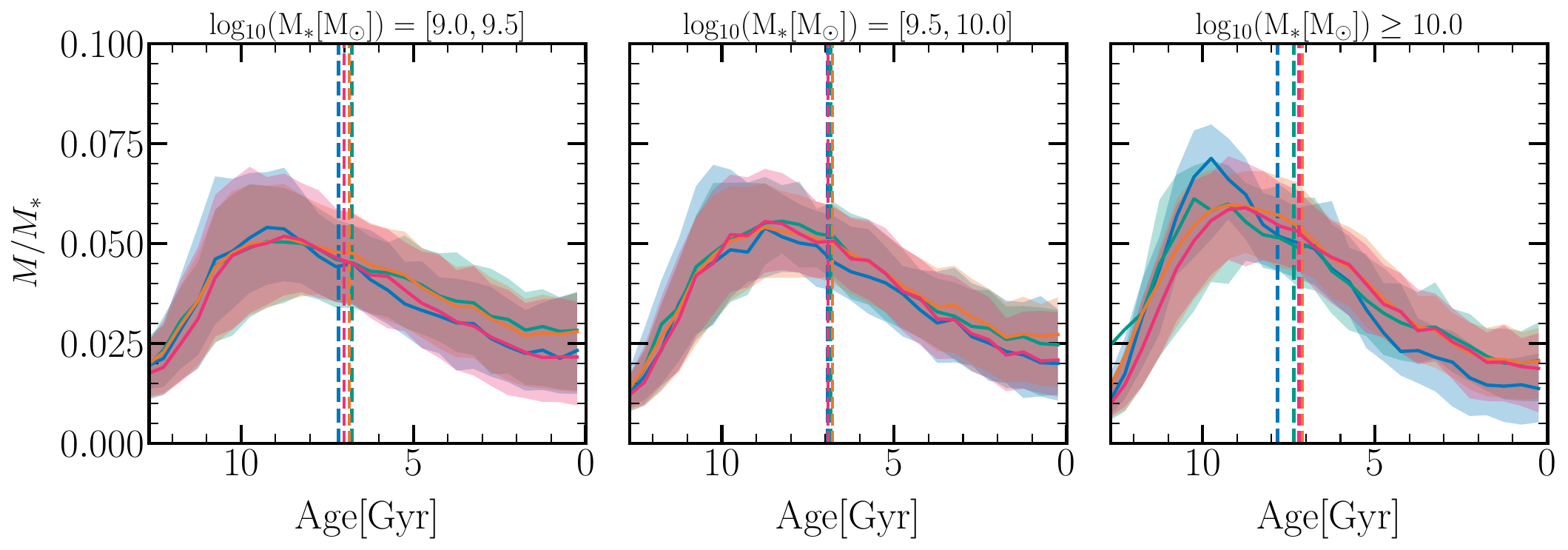} \\
    \includegraphics[width=2\columnwidth]{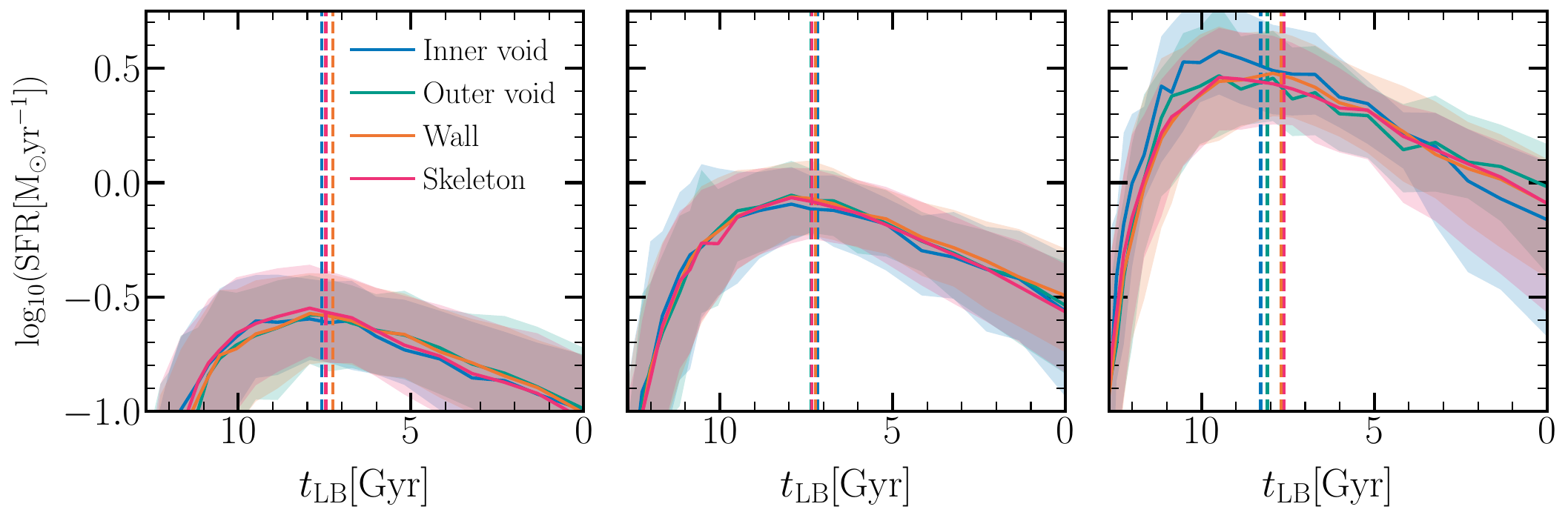} \\
    \caption{The median mass-weighted age distributions of the stellar population (top panels) and the star formation rates (SFRs) as a function of lookback time (bottom panels) for galaxies in the subsamples from different regions and inner galaxies and divided into three stellar mass bins. The shaded region represents the $25^{\rm th}$ and $75^{\rm th}$ percentiles. The vertical dashed lines represent the median of the stellar population age (top panels) and the galaxy formation time (bottom panels) for each subsample.}
    \label{fig:agehistogram}

\end{figure*}




\begin{figure*}

	\includegraphics[width=2\columnwidth]{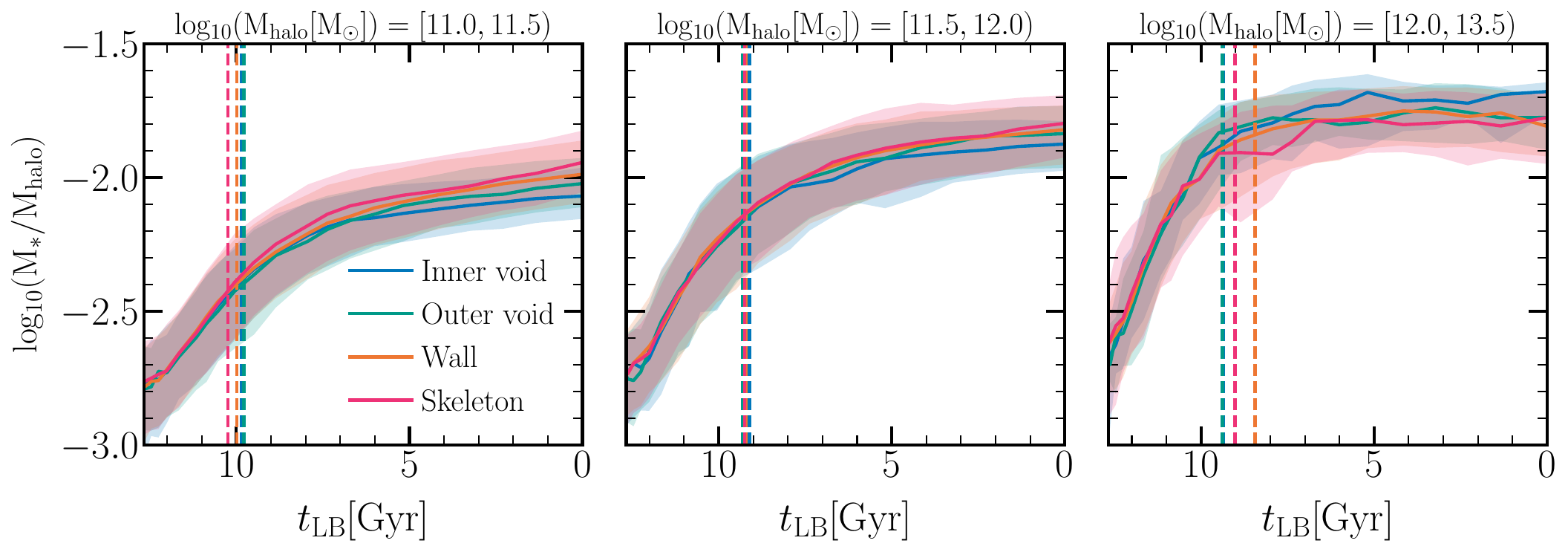}
	\caption{  The median evolution of the stellar mass fraction in haloes divided into three halo mass bins and for the different subsamples with the same stellar mass distribution. Dashed lines represent the median formation time for each sample, defined as the lookback time at which the halo assembles 50 per cent of its final mass. There is a significant difference in the stellar mass fraction in haloes with $M_{\rm halo}<10^{12}\Msun$ that has been slightly increasing over the last $7$ Gyrs.}
	\label{fig:HaloMasstoMstar}
\end{figure*}

In this section, we explore the evolution of galaxies and their host haloes using the same subsamples that have a similar stellar mass distribution as inner void galaxies and divide them into three stellar mass intervals.

To do this we define  $\tform$, the galaxy formation time, as the lookback time at which the main progenitor has assembled $50$ per cent of its final stellar mass ($z=0$)\footnote{ We caution the reader that this computation is conducted with the public \EAGLE database and has just 29 outputs in time. Although we interpolate the formation time, this might be noisy.}.
Figure \ref{fig:galaxyassembly} illustrates the growth in stellar mass relative to the final stellar mass at $z=0$. Low mass galaxies ($M_{*}=10^{[9,9.5)}\Msun$)   seem to form earlier (median $\tform=7.6_{-1.3}^{+1.2} $ Gyr, vertical dashed blued lines) than their analogues in denser environments (median  and  $25^{\rm th}-75^{\rm th}$ percentiles $\tform=7.2_{-1.2}^{+1.4}$ Gyr for wall galaxies, vertical magenta dashed lines). Although the scatter is similar to the difference in time between two consecutive outputs. This disparity is more obvious ($M_{*}\geq 10^{10}\Msun$), with inner void galaxies forming at $\tform=8.3_{-0.7}^{+0.9} $ Gyr and skeleton galaxies forming at $\tform=7.6\pm1 $ Gyr exhibiting the largest difference.    There is no discernible difference in the assembly of galaxies with intermediate stellar masses  ($M_{*}=10^{[9.5,10)}\Msun$) that have $\tform=7.2_{-1.2}^{+1.6} $  for inner void galaxies and $\tform=7.3_{-1.1}  ^{+1.1} $ Gyr for skeleton galaxies.

The top panel of Fig~\ref{fig:agehistogram} displays the fraction of the stellar mass formed at a given time for galaxies in the different subsamples and stellar mass bins. The median age of the stellar population is represented by the vertical dashed lines.  The stellar population of inner void galaxies with $M_{*}\geq 10^{10}\Msun$ is found to be older than that of galaxies living in other environments. However, we find no substantial difference in the stellar population age in galaxies with intermediate stellar masses as a function of the environment ($10^{[9.5,10)}\Msun$) and only a weak tendency for the lowest stellar mass galaxies.  This is also evident in the bottom panel of Fig~\ref{fig:agehistogram} for massive galaxies, where the star formation rate (SFR) is shown as a function of lookback time. Massive galaxies in inner voids are more star formation active at early times than massive galaxies in other environments, with the peak of star formation activity occurring at the same time as the peak of the distribution of the stellar population age, as expected. At later times, the star formation rates in void galaxies decline slightly faster than in other environments. In comparison,  when the environment changes, we find no significant difference in the star formation histories of intermediate and low stellar mass galaxies.


To conclude this section, we look at the evolution of the stellar mass fractions in Fig.~\ref{fig:HaloMasstoMstar} for three halo mass bins: $M_{\rm halo}=10^{[11,11.5)}\Msun$, $M_{\rm halo}=10^{[11.5,12)}\Msun$ and $M_{\rm halo}=10^ {[12,13.5]}\Msun$ and for the subsamples with a similar stellar mass distribution as the one of inner void galaxies.  The figure shows that the stellar mass fractions increase rapidly at earlier times and then flatten out later after the halo has already formed. This behaviour occurs for all halo mass bins. However, haloes with  $M_{\rm halo}<10^{12}\Msun$  in inner voids present lower stellar mass fractions ($0.85\times 10^{-2} \pm 1.5\times 10^{-4}$ and $1.33 \times 10^{-2} \pm 3.8\times 10^{-4}$ respectively at $z=0$) than haloes in other environments  ($1.13\times 10^{-2} \pm 1.8\times 10^{-4}$ and $1.59\times 10^{-2} \pm 9.1\times 10^{-4}$, respectively, for skeleton galaxies). This difference has been slightly increasing over the last $7$ Gyrs of evolution. This difference begins after the formation of inner void haloes, with the halo formation time ($t_{\rm form, h}$) defined as the interpolated lookback time at which a halo assembles 50 per cent of its final mass. Haloes with $M_{\rm halo}<10^{12}\Msun$) in inner voids  ($t_{\rm form, h}= 9.9_{-1}^{+0.8} \,\& \, 9.1_{-1.4}^{+1.2}$ Gyrs, respectively) seem to form later than those in other environments ($t_{\rm form, h}= 10.25_{-1.1}^{+0.7} \,\& \, 9.2_{-1.2}^{+1.0}$ Gyrs, respectively for skeleton haloes). Our findings are compatible with the results of \cite{alfaro2020} using the \textsc{TNG300} simulation \citep{nelson2018} and identifying voids via Voronoi tessellation of the galaxy catalogues \citep{ruiz2015a} although the halo masses that they explored are higher than this study.

In contrast, massive haloes in voids (inner regions) have slightly higher stellar mass fractions ($2.3\times 10^{-2} \pm 14.4\times 10^{-4}$) than those from haloes in the other subsamples ($1.7\times 10^{-2} \pm 17.1\times 10^{-4}$  for skeleton haloes). However, the jackknife errors are high.  Calculating the halo formation time for these haloes, we find that the halo formation time of the inner void massive haloes ($t_{\rm form, h}= 9.4_{-1.5}^{+0.6}$ Gyrs)  is larger than their massive counterparts in denser regions ($t_{\rm form, h}= 8.9_{-2.0}^{+0.9}$ Gyrs).  Although this difference is small, it might suggest that haloes in a denser environment continue growing, while haloes in inner voids regions, hardly accrete more matter at later times.

\section{Discussion}
\label{sec:discussion}
In this section, we bring together all our results to discuss what we can conclude about the effects of the large-scale environment on central galaxies by controlling the stellar mass.  The section concludes with an outlook on planned future work.

\subsection{What are the effects of the large-scale environment on massive central galaxies ($M_{*}>10^{10.25}\Msun$)?}

In section \ref{sec:galprop}, we found that massive galaxies in inner voids exhibit  considerable variance in their gas-phase metallicities and metallicity distributions. Both properties  are connected with the galaxies assembly history and have close relationships to the environment. Higher metallicities and negative metallicity gradients in massive galaxies are consistent with the fact that inner void galaxies are more isolated than those in denser settings. From Fig~\ref{fig:galaxyassembly} \& Fig.~\ref{fig:agehistogram}, we can observe that inner void galaxies were more star-forming active and contributed more gas to the cosmic web in early epochs compared to later epochs. In the right panel of Fig.~\ref{fig:discs}, the age of the youngest population of inner void galaxies are often older than those residing in denser environments confirming this. Once massive inner void galaxies formed, they possibly ceased accreting gas from the cosmic web, resulting in a decrease in star formation activity as seen in Fig.~\ref{fig:agehistogram}. Additionally, they have fewer mergers and interactions at later epoch (25 per cent of massive galaxies have at least one interaction), which might enrich the  gas and contribute to enhance SF activity and mix chemical elements, flattening the metallicity gradients. We could not identify  any significant differences in the SF activity and hydrogen gas fraction in inner void galaxies and their counterparts other environments, with the exception of outer void galaxies, which seem to be more SF active.  Upon investigating the SF activity in massive outer void galaxies, we found that a significant fraction (above of 90 percent) of them reside in voids with sizes smaller than $10$ pMpc.  \cite{paillas2017}
found that AGN feedback pollutes small voids with hot, metal-rich gas in EAGLE. This is consistent with the fact that massive galaxies in outer void regions have greater metallicities than in denser surroundings. \cite{deRossi2017} studied the impact of AGN feedback on the metallicity-stellar mass relation and found that AGN feedback were effective in expelling metal-rich gas in haloes with $M_{\rm halo} >10^{11.5}\Msun$ ($M_{*}=10^{10}\Msun$).
In contrast, we found a lower fraction of quenched massive galaxies in inner voids than those in other subsamples, which is compatible with their  past star formation activity. But because we did not find differences in the SF activity and hydrogen fractions, it is possible that the mechanism is related to the merger histories as suggested by \cite{ricciardelli2014} which noticed that the SF activity was not so different between galaxies in inner voids and galaxies  in other environments when considering only star-forming galaxies. Indeed,  we did observe that galaxies that experienced at least one merger ($\sim 50/25$ per cent for major/minor mergers in inset plots from Fig.~\ref{fig:mergersmstar}) had later mergers  (Fig.~\ref{fig:lastmergers_ssm}) although there was not significant difference in the merger histories. This may suggest  that  AGN feedback is the primary factor driving the disparities between massive galaxies. The lower efficiency of AGN feedback in inner voids could be related to the fact
that the mass of the BHs in the haloes is less massive than the rest of the subsamples in this stellar mass range (not shown here, Rosas-Guevara et al. in preparation). This may be in disagreement with \cite{habouzit2020}, who observed no difference in the BH mass-stellar mass relation at $z=0$ between BHs residing in underdense areas and the total population of BHs in the \textsc{HorizonAGN} simulation. The authors also used void galaxies and compared them to the rest of the simulation  over  two stellar mass bins ($M_{*}=10^{9-10}\Msun$ and $>10^{10}\Msun$), however, with satellites. Note that the BH mass-stellar mass relation is also different in both simulations \citep{habouzit2021}.


\subsection{What are the effects of the large-scale environment on low-mass galaxies ($M_{*}<10^{9.75}\Msun$)?}

In section \ref{sec:galprop}, we found that low-mass galaxies in inner voids show lower gas-phase metallicities on average (Fig.~\ref{fig:starmass-metallicidad}), with a higher fraction of negative metallicities than the rest of the subsamples,  albeit with a smaller difference (Fig.~\ref{fig:discs}).  In addition, low-mass galaxies in inner void regions have a slightly higher sSFR than the rest of the subsamples, but with an equivalent hydrogen gas fraction. These findings are compatible with the scenario where low-mass galaxies evolve slightly more slowly over time. This is also seen in \cite{habouzit2020} using \textsc{HorizonAGN} with different void finders.
However, based on  Fig~\ref{fig:galaxyassembly} \& Fig.~\ref{fig:agehistogram}, we cannot distinguish  between inner void galaxies and those in other cosmic web environments in terms of their star formation histories.  There is a minor trend  for inner void galaxies to form  ($\tform=7.6$ Gyr) earlier than those in other environments ($\tform=7.2$ Gyrs) but with a large scatter (more than 1 Gyr). To understand the low-metallicity gas in galaxies with the same stellar mass as in denser environments, it is necessary to see the merger histories. In this respect, we did not see a significant difference in the merger histories except for a small fraction of galaxies that experienced a major merger ($\approx 20$ per cent, see inset plots of Fig.~\ref{fig:mergersmstar}) where experienced a major merger earlier in time than their counterparts living in a denser environment.

Another possible explanation for the low metallicity of inner void galaxies with low stellar mass, is that the nature of the infalling gas into the host halo at earlier times could be different between inner void regions and denser regions, and this is consistent with the host haloes of inner void galaxies formed later in comparison with host haloes in other regions (see Fig.~\ref{fig:HaloMasstoMstar}).

\subsection{What are the effects of the large-scale environment on intermediate central galaxies ($M_{*}=10^{9.75,10.25}\Msun$)? }

In the case of intermediate central galaxies, we found that this stellar mass range  is crucial  due to the changing patterns in galaxy properties. Within this stellar mass range, galaxies in inner voids present lower hydrogen fractions. This has been reported by \cite{dominguez2022} who used observed CO emission lines for these galaxies and compared molecular and atomic gas between galaxies in voids and filaments. The authors find that inner void galaxies present less hydrogen gas than their counterparts in denser environments. In addition,  intermediate stellar mass galaxies present higher gas-phase metallicities, lower SF activity, and a higher fraction of quenched galaxies.  As previously mentioned, AGN feedback becomes effective at haloes above $M_{\rm halo}=10^{11.5}\Msun$ corresponding to $M_{*}=10^{10}\Msun$ \citep{deRossi2017,habouzit2021}.  In addition, we also found that the BHs in this stellar mass range, become more massive in inner voids ( mean $\mbh=10^{6.5}\Msun$) than those in other dense environments (mean $\mbh=10^{6.2}\Msun$ in skeleton galaxies) (Rosas-Guevara et al. in preparation).  In addition, the fact that 70 per cent of quenched galaxies reside in voids smaller than $10$ pMpc suggests that the size of the voids is a crucial factor in determining these disparities. This is consistent with the findings of \cite{paillas2017} that showed that galaxy feedback processes can contaminate voids with hot, metal-rich gas, specially voids with $\Rvoid<10$ pMpc.

\subsection{Outlook}
One caveat in our analysis is that the galaxy subsamples with identical stellar mass distribution have a distinct population of host haloes, which might indicate that all of the variances in the galaxy properties could be driven by the distinct populations of haloes in the distinct environments. When we employ galaxy subsamples with the same halo distribution as the one found in inner void galaxies, the differences seen in the key galaxy properties remain, indicating that the differences are not only caused by different populations of haloes. The drawback of this approach is that our subsamples will have different stellar mass distributions along with environments.  To examine the impact of a large-scale environment, the halo and stellar mass distributions of the distinct environments should be matched. However, imposing this condition reduces our samples to smaller sizes and mass ranges, both of which are already limited. In future investigations, it may be possible to investigate this in larger cosmological hydrodynamical simulations with at least the same resolution as the current one, but with a broader spectrum of cosmic voids, including those that are larger than the ones observed in SDSS.

Our findings suggest that the void dimensions have an impact on the properties of the galaxies that inhabit them.  For example, the modest excess of quenched galaxies in inner voids compared to skeleton galaxies  can be associated with galaxies inhabiting voids with limited dimensions. This hints that the history of smaller voids  would be different.  Indeed, the population of voids can be subdivided based on the environment dynamics. \cite{sheth2004}, split voids into two categories: void-in-void and void-in-cloud. The first kind of voids relates to those that are embedded in an underdense environment with expanding walls, whereas the second type corresponds to those that are immersed in an overdense environment with contracting walls.
This duality was initially identified by \cite{ceccarelli2013}, who classified voids as R-type or S-type based on their integrated density contrast profile.
Small voids are often surrounded by overdense walls, but big voids frequently have underdense surrounds and continuously increasing integrated density profiles. As void size grows, the fraction of S-type voids decreases. \cite{ruiz2015a} found that the nonlinear dynamics of haloes in the inner regions of voids depends on their kind. \cite{rodriguez2022} investigate how this disparity in void qualities influence the formation, evolution of the haloes and their hosted galaxies using zoom-in cosmological hydrodynamical simulations. However, this should be studied using the next generation of cosmological hydrodynamical simulations of galaxy formation.

One aspect of our analysis that remains unresolved is to disentangle the processes that are fostered in different environments that drive the variations in the galaxy properties found at $z=0$, such as gas-phase metallicities and their gradients. These properties are thought to encapsulate information from the recent interactions and the nature of the infalling gas into the haloes.  Indeed, we do find hints that the merger history of galaxies may play a role \citep{tissera2021}, but the study the properties of the infalling gas into the haloes in diverse environments is beyond the scope of this paper.

Another mechanism that have not been explored in this paper is the impact of AGN feedback. It is believed that black holes could grow by secular evolution in underdense regions and then affect the star-forming activity in the galaxy differently than the environment (see \citealt{habouzit2020}). Indeed, the influence of the large-scale environment on the feeding and fuelling of BHs and their effects on their host galaxies needs to be understood. In a forthcoming paper, we will investigate this by exploring the BH properties in cosmic voids.

\section{Summary}
\label{sec:summary}
We analyse the properties of galaxies as a function of the void-centric distance using the largest $\Lambda$CDM  cosmological hydrodynamical simulation from the \EAGLE project \citep  {schaye2015,crain2015,furlong2015a}. 
Our research focuses on central galaxies with a stellar-mass larger than $10^9 \Msun$, yielding, 7400 central galaxies. To identify voids in the simulation, we use the void catalogue constructed in \cite{paillas2017} with a spherical under-density finder \citep{padilla2005}. 
Based on the void-centric distance, we defined four parent samples of galaxies: Inner void, outer void, wall and skeleton galaxies.

Our findings are as follows:
\begin{itemize}
\item We find that the stellar mass distribution of galaxies in inner and outer voids are biased towards low-stellar mass galaxies  (see Fig.~\ref{fig:gmf}),  consistent with observations and cosmological N-body simulations \citep {croton2005,moorman2016,ricciardelli2014}.
\item We also study the halo mass-stellar mass relation, finding that galaxies in voids have smaller stellar-mass fractions than galaxies in denser regions for halo masses of $M_{\rm halo}=10^{11,12}\Msun$ (see Fig.~\ref{fig:gmf}). This is in agreement with other cosmological hydrodynamical simulations \citep[e.g.][]{alfaro2020,habouzit2020}.

\item The fraction of star-forming galaxies and galaxies with no null hydrogen gas fraction decreases with increasing void-centric distance (see Fig.~\ref{fig:fractionvoids}), which is in agreement with observations from SDSS DR4 \citep[e.g.][] {ricciardelli2014}.

\end{itemize}

To study the consequences of the large-scale environment on galaxies, we concentrated on galaxy subsamples of different environments with the same stellar mass distribution as the galaxies located in the inner void region, and in appendix \ref{app:halosame}, we show these properties with subsamples with the same halo mass distributions. Our findings are as follows:

\begin{itemize}
\item  For the galaxy subsamples, we explore the halo mass-stellar mass relations. The lower stellar mass content seen in inner void galaxies compared with galaxies in other environments is preserved for a given halo mass with $M_{\rm halo}=10^{(11,12)}\Msun$ (see Fig.~\ref{fig:gmf_samestellarmass}), but the reverse trend is shown for the most massive haloes ($M_{\rm halo}\geq10^{12}\Msun$).

\item  We find that galaxies in inner voids with $M_{*}=10^{9.0,9.5}\Msun$ tend to have slightly higher star-formation activity and similar HI gas fractions than their counterparts in denser environments (see Fig.~\ref{fig:samestellarmass}). For the most massive galaxies, there is no significant difference in SF activity and HI gas fractions with the environment, except for the outer void galaxies, which have the highest SF activity and HI gas fractions. However, the jackknife errors associated with these differences are large.


\item We examine the fraction of quenched galaxies (sSFR$<10^{11.5}\rm yr^{-1}$).  Overall,  this fraction rises with increasing void-centric distance from $0.04$ in the inner void regions to  $0.13$ in the skeleton regions  (see Fig.~\ref{fig:quenchedfraction}). This suggests that quenching mechanisms that are more efficient in denser regions, such as ram pressure, take place when the void-centric distance is increased. In addition, the level of current star-formation activity in inner voids seems to be modulated by other mechanisms, such as the lack of replenishing gas and/or mergers and interactions.

\item We explore the stellar mass-gas phase metallicity relation, using oxygen abundances of the star-forming gas as a proxy  (see Fig.~\ref{fig:starmass-metallicidad}).  Galaxies with $M_{*}=10^{[9.5,10]}\Msun$ have a flat relation, with inner void galaxies having the lowest gas-phase metallicities. The stellar mass-gas phase metallicity relation increases with stellar mass for galaxies in inner voids with $M_{*}>10^{10}\Msun$ compared to galaxies in denser environments where the relation becomes flatter.



\item In terms of morphology, the fraction of galaxies with a dominant disc component increases with stellar mass regardless of the large-scale environment, except for skeleton galaxies (see Fig.~\ref{fig:morphclass}) with 29 per cent of galaxies in inner voids and 30 per cent in skeleton galaxies having a dominant disc. However, there is a higher fraction of massive disc galaxies ($ 68\%$ for $M_{*}\geq 10^{10.2}\Msun$)  in inner void regions than in other regions ($56\%$ and $58\%$ in skeleton and wall galaxies).

\item We analyse the gas-phase metallicity gradients and the young stellar populations (see Fig.~\ref{fig:discs}) of galaxies with resolved discs, and divide them into two stellar mass bins: $>10^{10}\Msun$ and $\leq10^{10}\Msun$. We find that galaxies are more likely to have negative metallicity gradients in inner voids (80 per cent of the galaxies with negative gradients and a median of $-0.009$ with jackknife errors of $0.008\, \rm dex \,kpc^{-1}$)  than in denser environments (70 per cent of skeleton galaxies have negative gradients and a median of $-0.004$ with jackknife errors of $0.001 \, \rm dex\, kpc^{-1}$). This difference is observed for both stellar mass bins, with a stronger signal for the highest stellar mass bins. We also find that the youngest stellar population in inner voids are older than in the other regions overall.

\end{itemize}

To gain better understanding of the differences found between inner void galaxies and those denser environments, we track the evolution of galaxies in inner voids and compare them with subsamples from other environments that have  the same stellar mass distribution and are divided into three stellar mass bins: $10^{[9.,9.5)}\Msun$,$10^{[9.5,10)}\Msun$ and $\geq10^{10}\Msun$.
The following are our findings:

\begin{itemize}

\item We find a modest difference in the merger histories of galaxies as a function of their location in the cosmic web (see Fig.~\ref{fig:mergersmstar}). We find that the number of major mergers increases with stellar mass and galaxies with $M_{*}=10^{9.5,10,5}\Msun$ in inner voids present the lowest number of major mergers and the higher minor mergers. Whereas, galaxies in the skeleton with larger stellar masses present the highest number of minor mergers. For the lowest stellar mass galaxies, there is no significant difference.

\item We also investigate the lookback time distribution of the last major/minor mergers (see Fig.~\ref{fig:lastmergers_ssm}). For low-stellar mass galaxies ($10^{[9.,9.5)}\Msun$, the last major merger seems to happen  earlier  in voids, whereas their last minor merger occurred later than galaxies living in the other environments.  For intermediate and massive galaxies ($10^{[9.5,10)}\Msun$ and $\geq10^{10}\Msun$), their last  major merger happens later in inner voids  than those in denser environments with  more diversity in the major merger histories.  A similar trend is found in minor mergers: void galaxies undergo their last minor mergers later  than  galaxies in denser environments, except for galaxies with $M_{*}=10^{[9.5,10)}\Msun$.


\item Lower stellar mass fractions ($0.85\times 10^{-2} \pm 1.5\times 10^{-4}$ and $1.33 \times 10^{-2} \pm 3.8\times 10^{-4}$) in low and intermediate halo masses ($M_{\rm halo}=10^{[11,11.5)\Msun}$  and $10^{[11.5,12)\Msun}$, respectively) in inner voids than those in the other environments ($1.13\times 10^{-2} \pm 1.8\times 10^{-4}$ and $1.59\times 10^{-2} \pm 9.1\times 10^{-4}$) are detected. This difference has started since $7$ Gyrs ago. In contrast, massive haloes ($10^{[12,13.5)\Msun}$)  in inner void environments have similar stellar mass fractions ($2.3\times 10^{-2} \pm 14.4\times 10^{-4}$) to those living in other environments ($1.7\times 10^{-2} \pm 17.1\times 10^{-4}$ for skeleton haloes) since early times.

\end{itemize}

We find that the key properties of galaxies in a denser, large-scale environment with the same stellar mass distribution as those galaxies in inner voids differ from those galaxies in the inner part of voids. While these disparities are small, they become enhanced when the stellar mass is not controlled, as shown in observations \citep[i.e.][]{ricciardelli2014} and simulations \citep[i.e.][]{croton2005}.

Overall, our results show how large-scale environments could have an effect on the evolution of central galaxies and that this could have imprinted the properties of the galaxies.

\section*{Acknowledgements}
The authors thank the referee for the constructive comments of the manuscript that improved the clarity of the paper. The authors also thank Jesus Dominguez-Gomez and Agustin Rodriguez-Medrano for the useful comments of the manuscript. YRG acknowledges the support of the
“Juan de la Cierva Incorporation” fellowship (ĲC2019-041131-I) and the European Research Council through grant number ERC-StG/716151. PBT acknowledges partial funding by Fondecyt 1200703/2020
(ANID) and ANID Basal Project FB210003. NP gratefully acknowledges support by Fondecyt Regular 1191813, the ANID BASAL projects ACE210002 and FB210003. CL has received funding from the ARC Centre of
Excellence for All Sky Astrophysics in 3 Dimensions (ASTRO 3D), through project number CE170100013.
CL also thanks the MERAC Foundation for a Postdoctoral Research Award. This project has been supported partially by the European Union Horizon 2020 Research and Innovation Programme
under the Marie Sklodowska-Curie grant agreement No 734374.
We acknowledge the Virgo Consortium for making their simulation data available. The \EAGLE simulations were performed using the DiRAC-2 facility at Durham, managed by the ICC, and the PRACE facility Curie based in France at TGCC, CEA, Bruyeres-le-Chatel.
This work used the DiRAC Data Centric system at Durham University, operated by the Institute for Computational Cosmology on behalf of the STFC DiRAC HPC Facility (\url{www.dirac.ac.uk}). This equipment was funded by BIS National E-infrastructure capital grant ST/K00042X/1, STFC capital grant ST/H008519/1, and STFC DiRAC Operations grant ST/K003267/1 and Durham University. DiRAC is part of the National E-Infrastructure.  We thank contributors to {\sc SciPy} \footnote{http://www.scipy.org},{\sc
Matplotlib} \footnote{http://www.matplotlib.sourceforge.net}, and
the {\sc Python} programming language \footnote{http://www.python.org}


\section*{Data Availability}
The data from the \EAGLE simulations can be found on the website: \url{https://www.http://icc.dur.ac.uk/Eagle/database.php} \citep[see][]{mcAlpine2016}.




\appendix

\section{galaxy properties with the same halo distribution}
\label{app:halosame}
\begin{figure}
	\begin{tabular}{c}
	\includegraphics[width=1\columnwidth]{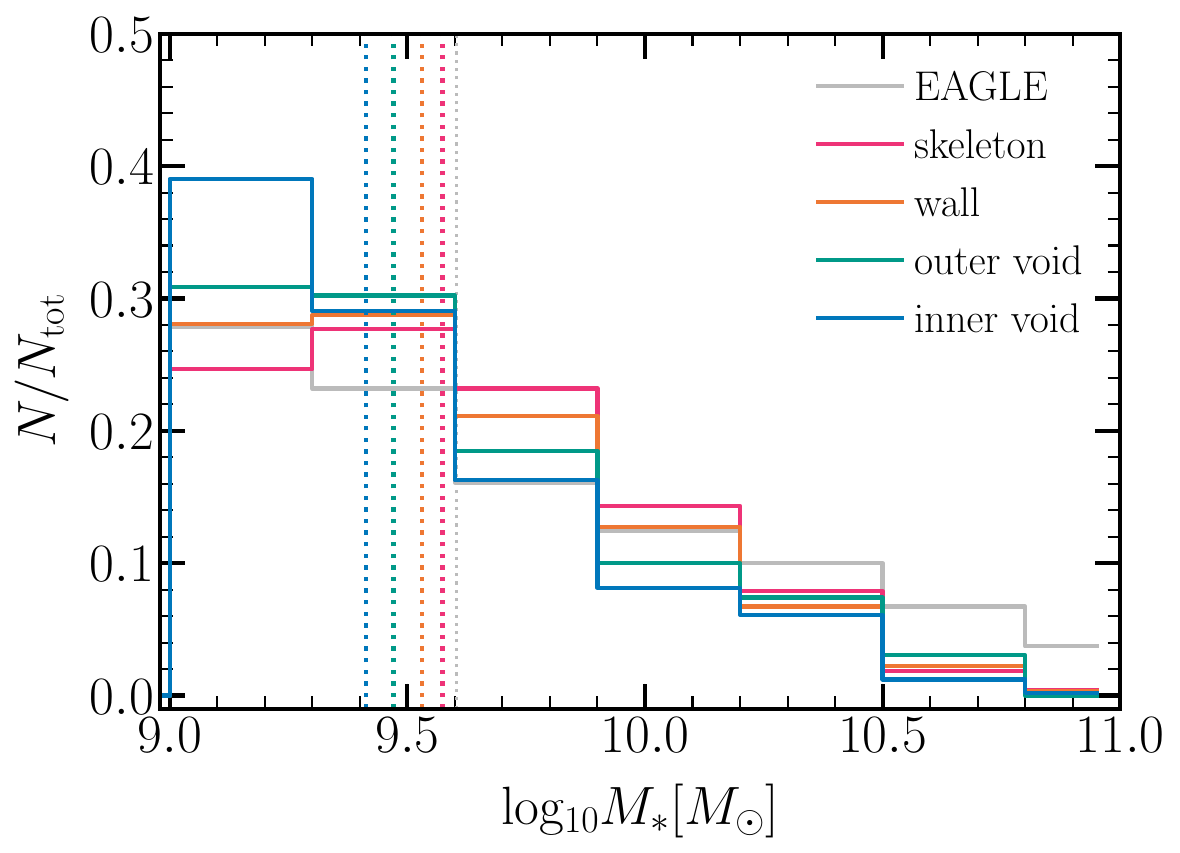} \\
	\includegraphics[width=1\columnwidth]{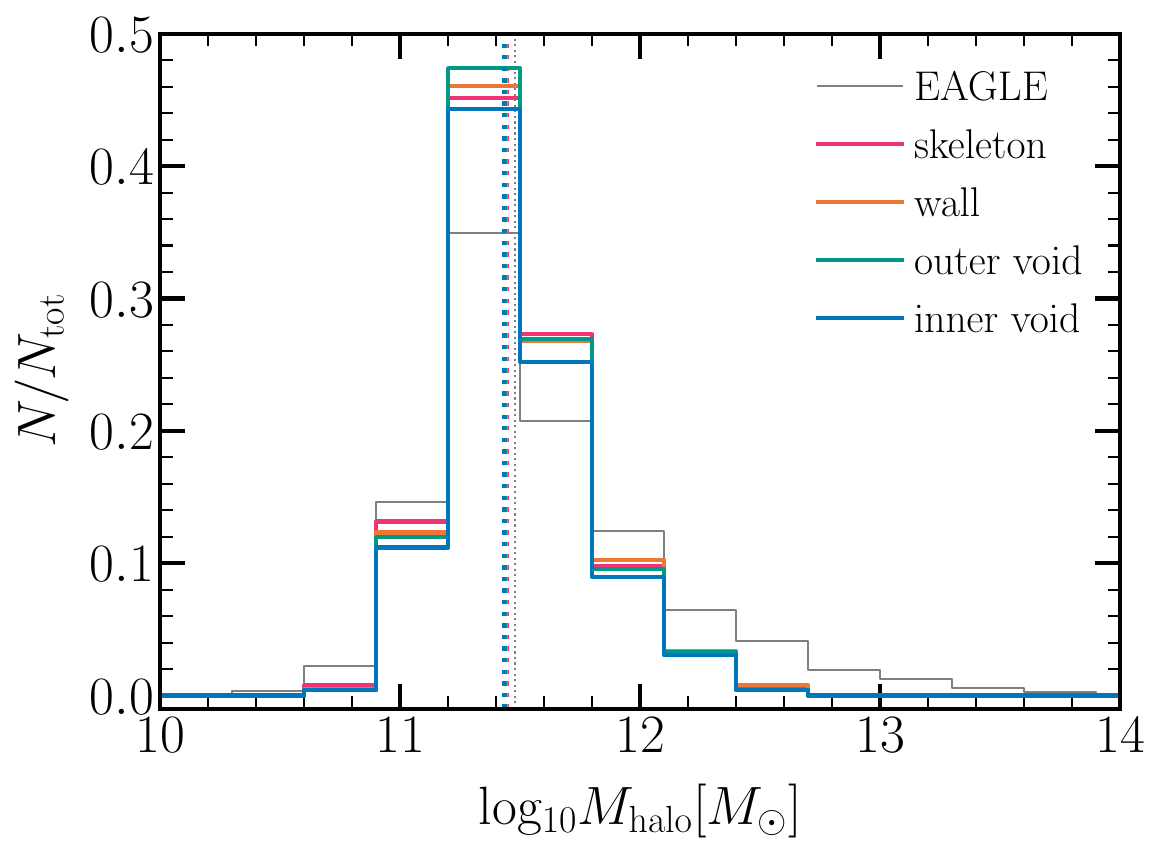}\\
	\includegraphics[width=1.0\columnwidth]{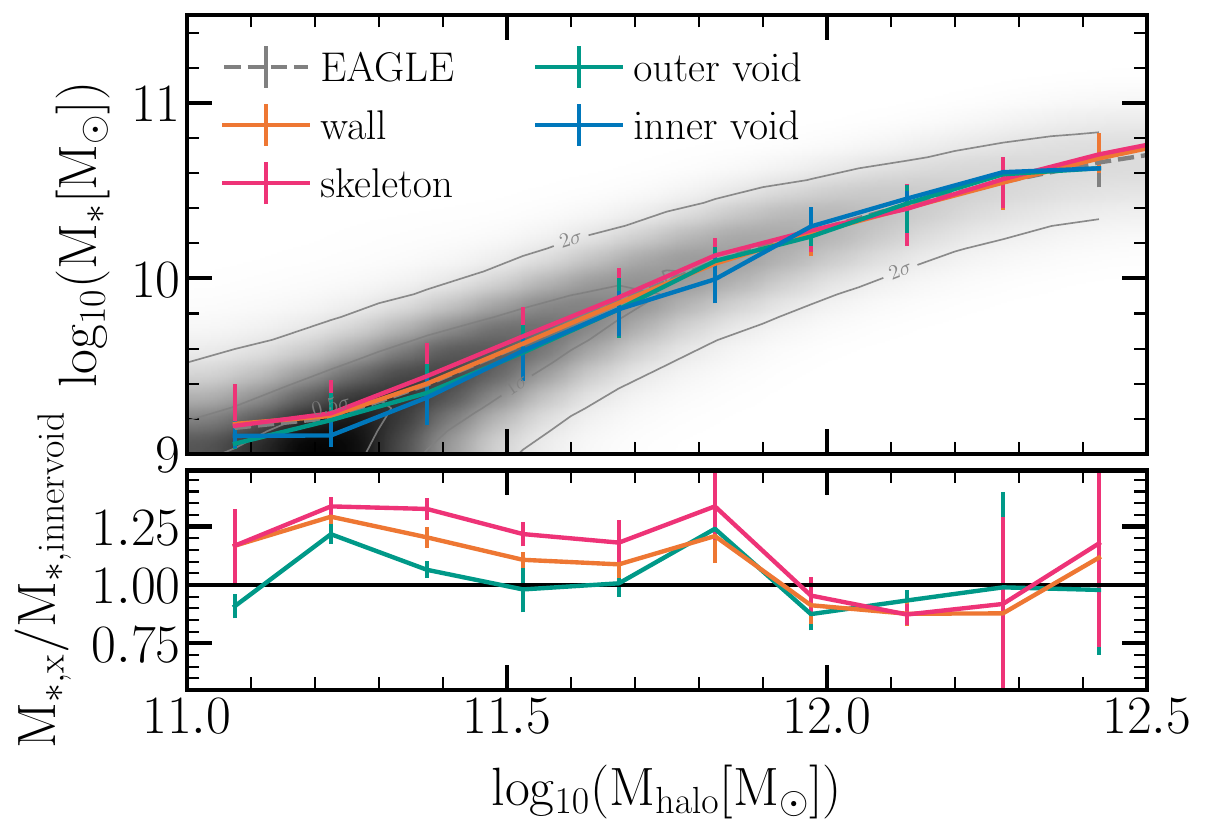} \\
	\end{tabular}
    \caption{ As in Fig~\ref{fig:gmf_samestellarmass} but using subsamples with the same halo mass distribution. The stellar mass (top panel) and halo mass (middle panel) distributions and the halo mass and stellar mass relation (bottom panel). Vertical lines represent the median of each distribution. Grey colour and the diffused density map and contours represent the distribution of all galaxies in the simulation. Error bars represent the $20^{\rm th}$ and $80^{\rm th}$ percentiles of each sample. The ratio between the stellar mass of each subsample and the median distribution of inner void galaxies for a given halo mass are shown in the bottom figure. Errorbars in the ratio corresponds to jackknife errors with $10$ subsamples. As for subsamples with the same stellar mass, inner void haloes host lower stellar mass galaxies in haloes with $M_{\rm halo}10^{[11-12.2]}\Msun$.}
    \label{fig:gmf_samehalomass}
\end{figure}

\begin{figure*}
	\begin{tabular}{cc}

	\includegraphics[width=1.1\columnwidth]{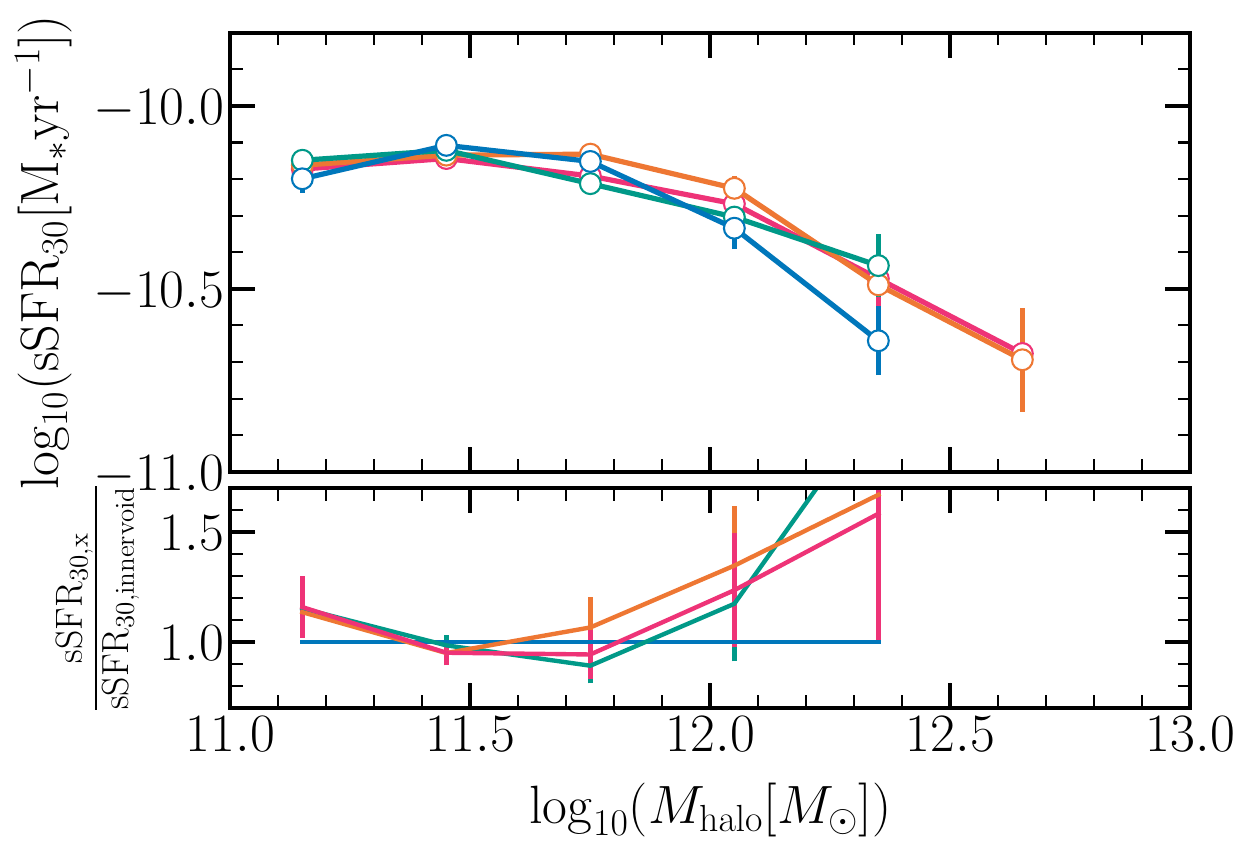} &
    \includegraphics[width=1\columnwidth]{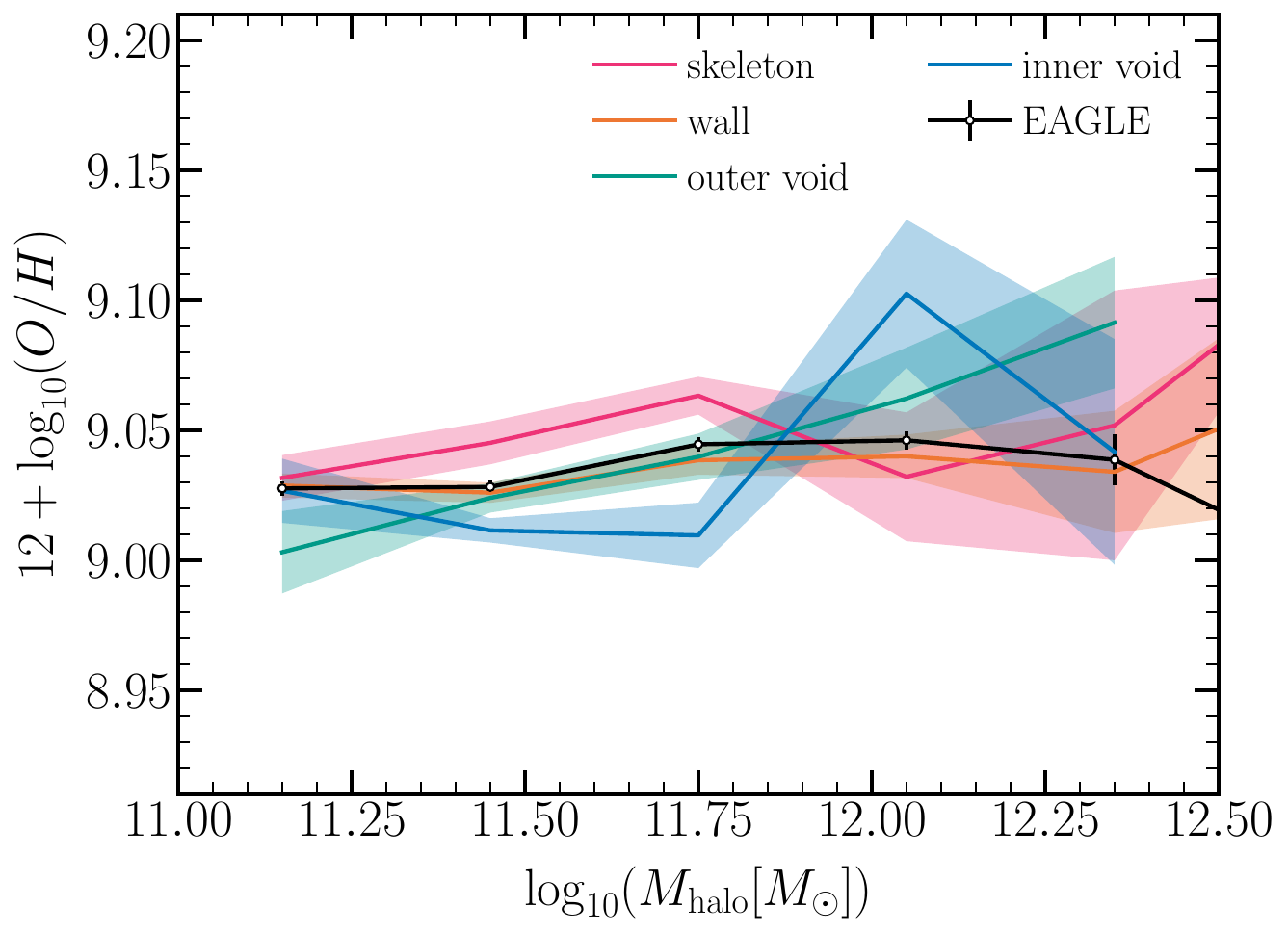}\\
    \includegraphics[width=1\columnwidth]{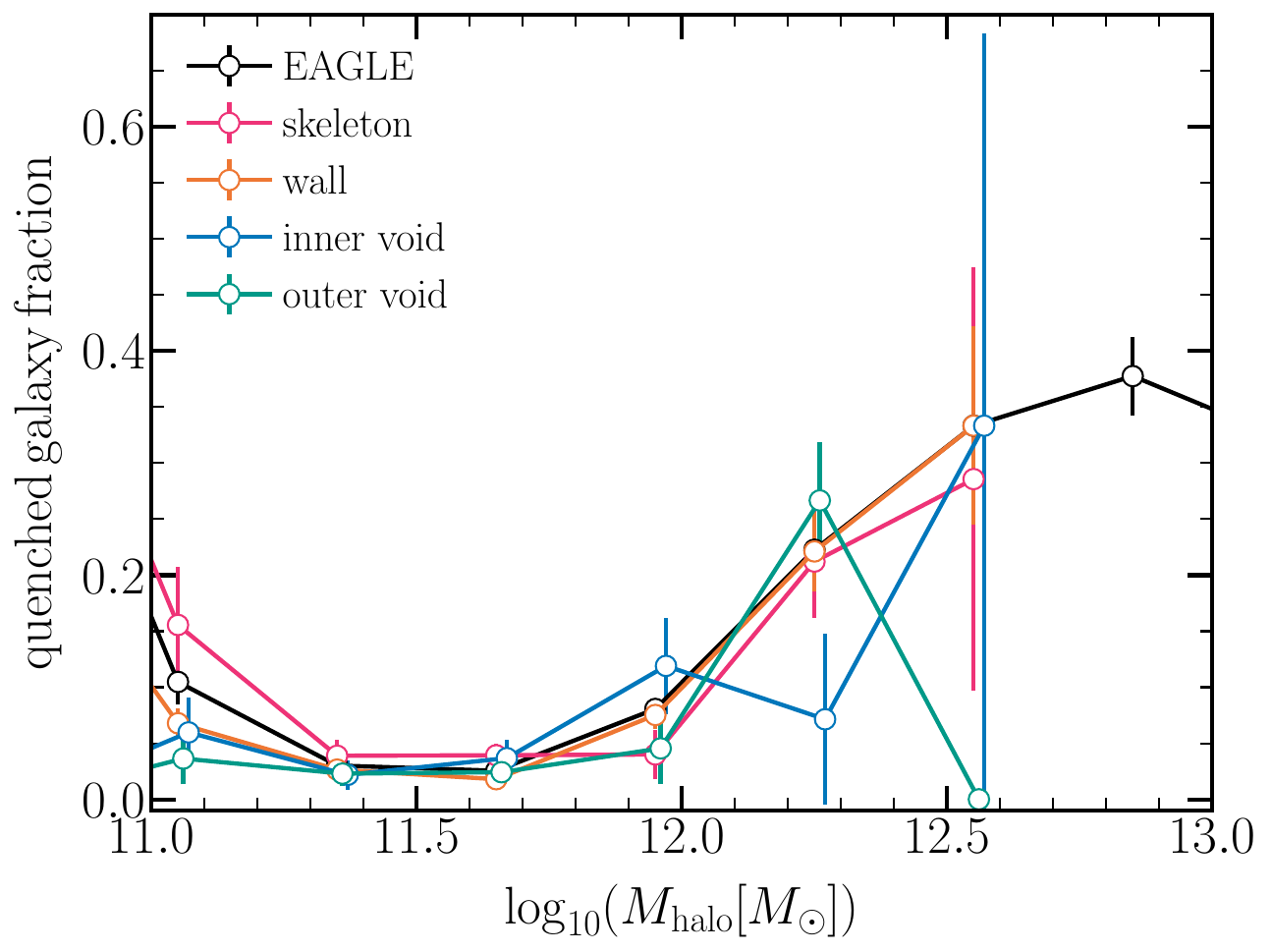}
   &

	\includegraphics[width=1\columnwidth]{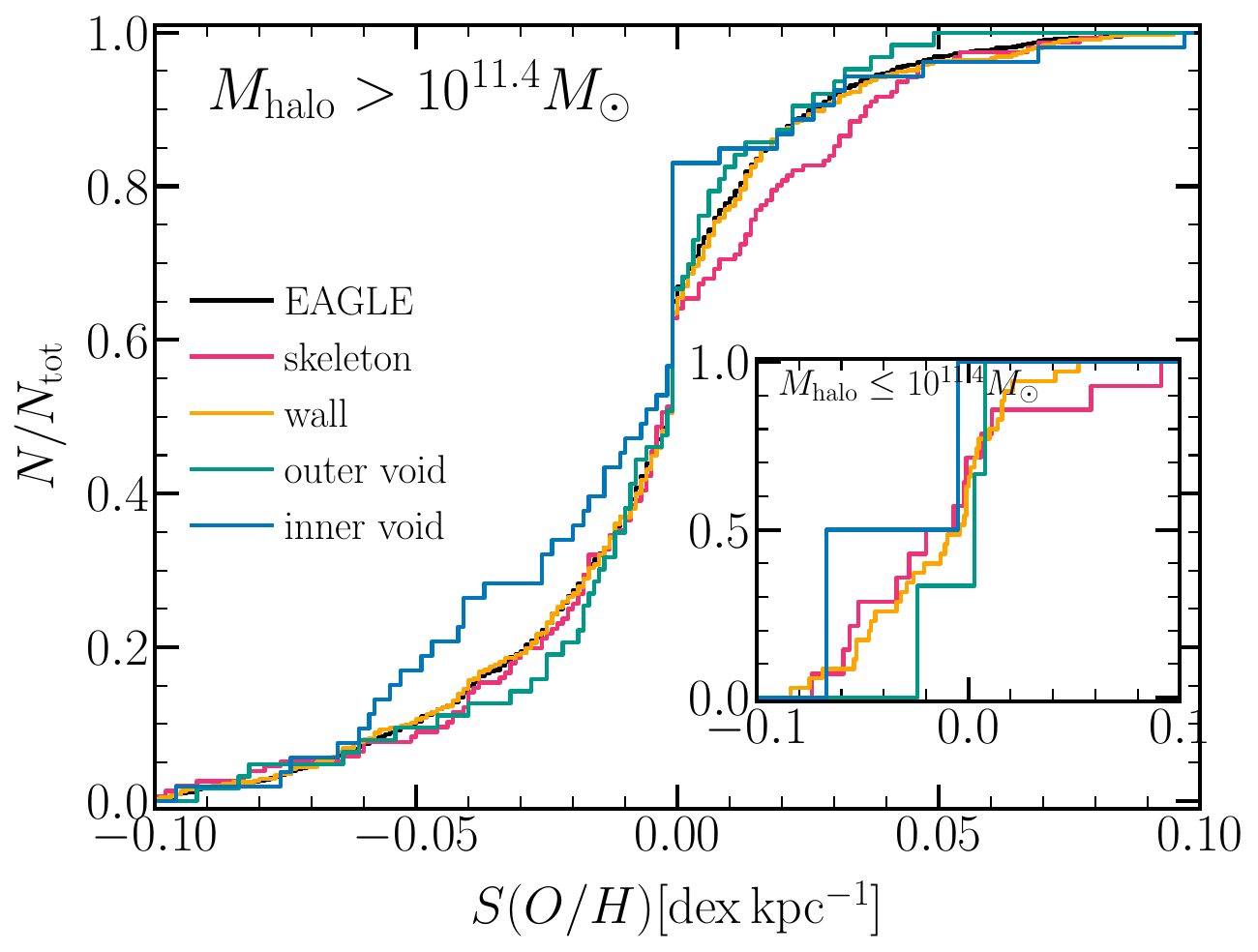} \\
	\end{tabular}
    \caption{Properties of galaxies in the subsamples with the same halo mass distribution, as the legend indicates. Errorbars and shaded regions correspond to jack knife errors using 10 subsamples. \textit{Top left panel:} The mean specific star formation rate (sSFR) as a function of halo mass. Coloured markers and solid lines represent the mean distributions. \textit{Top right  panel:} The mean relation between halo mass and the star-forming gas-phase oxygen abundance. The black solid lines and with circles represent the mean relation with entire simulation. \textit{Bottom left panel:} The fraction of galaxy that are quenched (sSFR$<10^{-11.5}\rm yr^{-1}$) as a function of halo mass. \textit{Bottom right panel:} The cumulative gas-phase oxygen abundance distribution in star-forming galaxies including  with halo masses $>10^{11.4}\Msun$ as is indicated in the legend and for stellar masses $\leq 10^{11.4}\Msun$ in the inset plot.}
    \label{fig:prop_samehalomass}
\end{figure*}

In this section, we present the main galaxy properties that have been shown in section \ref{sec:galprop} but using subsamples with the same halo mass distribution as the inner void samples instead of the same stellar mass distribution. The subsamples with the identical stellar mass distribution have a distinct population of haloes, as we found in section \ref{sec:galprop}. haloes in inner voids are more massive than those in denser environments (see Fig. \ref{fig:gmf_samestellarmass}). This might indicate that all of the variances in the galaxy properties could be driven by distinct populations of haloes.

To investigate this,  we should ideally take distributions of galaxies in various regions that have the same stellar mass and halo mass ratio distributions. Imposing this condition on our samples, on the other hand, drastically reduces the sample sizes.
However, we explore subsamples of the different environments that have the same halo distributions. The halo mass distribution, stellar mass distribution, and stellar mass as a function of halo mass are shown in Fig~\ref{fig:gmf_samehalomass} for these subsamples. We can observe that the differences found are preserved. Inner void galaxies have a stellar mass distribution biased to lower stellar masses, and they have a lower stellar mass content for a fixed halo mass in $M_{\rm halo}10^{[11-12.2]}\Msun$.

The sSFRs, quenched galaxy fractions, and gas-phase metallicities as a function of halo masses are depicted in Fig~\ref{fig:prop_samehalomass}.  With the exception of the sSFR-halo mass diagram, we detect discrepancies between inner void galaxies and those galaxies in denser environments, as in Figs.  ~\ref{fig:samestellarmass}, \ref{fig:quenchedfraction}. \ref{fig:starmass-metallicidad} and  \ref{fig:discs}, respectively.  We find that inner void galaxies with lower halo mass ($M_{\rm halo}\leq 10^{11.4}\Msun$)  have lower quenched galaxy fractions, lower gas-phase metallicity, and higher fractions of negative gas-phase metallicity gradients ($1$) than their analogues ($0.71$ for skeleton galaxies) in denser environments. The differences, however, are not as pronounced as they were in the subsamples with the same stellar mass distribution.

We find that galaxies in massive haloes ($M_{\rm halo}> 10^{11.4}\Msun$)  contain more gas-phase metallicity, with a slightly higher quenched fraction of massive galaxies than their halo counterparts in denser environments, with the exception of the highest halo mass bin, which has a similar fraction but with higher jackknife errors.
In addition, we find that inner void haloes have a higher fraction of negative gas-phase metallicity gradients ($0.83$) than their analogues in denser environments ($0.63$ for skeleton galaxies).It is worth noting that we observe comparable differences when we use the halo population of the subsamples with the same stellar mass distribution. The findings shown here may indicate that the differences seen are not entirely driven by the halo population in each environment.


\bsp	
\label{lastpage}
\end{document}